\documentclass[12pt,draftclsnofoot,a4paper,onecolumn]{IEEEtran}

\usepackage[latin1]{inputenc}
\usepackage[T1]{fontenc}
\usepackage{amssymb, amsmath, amsthm, latexsym}
\usepackage{dsfont}
\usepackage{graphicx}        
\usepackage{subfigure}                
\usepackage{epsfig}
\usepackage{nicefrac}
\usepackage{bm}
\usepackage{cite}

\DeclareGraphicsExtensions{.eps}

\newcommand{\ie}{\textit{i.e.,~}}%
\newcommand{\etal}{\textit{et al}.}

\begin{document}
%
\title{Construction of New Delay-Tolerant\\ Space-Time Codes}

\author{Mireille Sarkiss, Ghaya Rekaya-Ben Othman, Mohamed Oussama Damen\\ and Jean-Claude Belfiore}

\maketitle
\IEEEpeerreviewmaketitle

\begin{abstract}
Perfect Space-Time Codes (STC) are optimal codes in their original construction for Multiple Input Multiple Output (MIMO) systems. Based on Cyclic Division Algebras (CDA), they are full-rate, full-diversity codes, have Non-Vanishing Determinants (NVD) and hence achieve Diversity-Multiplexing Tradeoff (DMT). In addition, these codes have led to optimal distributed space-time codes when applied in cooperative networks under the assumption of perfect synchronization between relays. However, they loose their diversity when delays are introduced and thus are not delay-tolerant. In this paper, using the cyclic division algebras of perfect codes, we construct new codes that maintain the same properties as perfect codes in the synchronous case. Moreover, these codes preserve their full-diversity in asynchronous transmission.
\end{abstract}


\section{Introduction and Problem Statement}

\IEEEPARstart{D}{uring} the past decade, MIMO techniques have experienced a great interest in wireless communication systems. Using multiple antennas at the transmitter and the receiver provides high data rates and exploits the spatial diversity in order to fight channel fadings and hence improve the link reliability. Lately, cooperative diversity has emerged as a new form of spatial diversity via cooperation of multiple users in the wireless system \cite{Sandonaris:2003-I}. While preserving the same MIMO benefits, it counteracts the need of incorporating many antennas into a single terminal, especially in cellular systems and ad-hoc sensor networks, where it can be impractical for a mobile unit to carry multiple antennas due to its size, power and cost limitations.

In cooperative networks, users communicate cooperatively to transmit their information by using distributed antennas belonging to other independent terminals. This way, a virtual MIMO scheme is created, where a transmitter is also acting as a relay terminal, with or without some processing, assisting another transmitter to convey its messages to a destination. The cooperative schemes have been widely investigated by analyzing their performance through different cooperative protocols \cite{Sandonaris:2003-I, Laneman:2003, Laneman:2004}. These protocols fall essentially into two families: Amplify-and-Forward (AF) and Decode-and-Forward (DF). In order to achieve the cooperative diversity, space-time coding techniques of MIMO systems have also been applied yielding many designs of distributed space-time codes under the assumption of synchronized relay terminals \cite{Laneman:2003, Laneman:2004, Jing:2006}.

However, this \textit{a priori} condition on synchronization can be quite costly in terms of signaling and even hard to handle in relay networks \cite{Wei:2004, Li:2005}. Unlike conventional MIMO transmitter, equipped with one antenna array using one local oscillator, distributed antennas are dispersed on different terminals, each one with its local oscillator. Thus, they are not sharing the same timing reference, resulting in an asynchronous cooperative transmission.

On the other hand, in a synchronous transmission, the distributed STCs are constructed basically according to the rank and determinant criteria \cite{TAROKH:1998} and hence aim at achieving full diversity. Note that the rows of the codeword matrix represent the different relay terminals (antennas). So, when asynchronicity is evoked, delays are introduced between transmitted symbols from different distributed antennas shifting the matrix rows. This matrix misalignment can cause rank deficiency of the space-time code, and thus performance degradation.

Therefore, the codes previously designed are no more effective unless they tolerate asynchronicity. Furthermore, an efficient code design should satisfy the full-diversity order for any delay profile. This intends to guarantee full-rank codewords distance matrix \ie its rank equal to the number of involved relays, hence leading to the so-called delay-tolerant distributed space-time codes \cite{Li:2005}.

\section{Delay-Tolerant Distributed Space-Time Codes}

The first designs of such codes were presented by Li and Xia \cite{Li:2005} as full-diversity binary Space-Time Trellis Codes (STTC) based on the Hammons-El Gamal stacking construction, its generalization to Lu-Kumar multilevel space-time codes, and the extension of the latter codes for more diverse AM-PSK constellations \cite{Hammons:2005-1,Hammons:2005-2}. Systematic construction including the shortest STTC with minimum constraint length was also proposed in \cite{Shang:2006}, as well as some delay-tolerant short binary Space-Time Block Codes (STBC) \cite{Hammons:2006}. Recently, Damen and Hammons extended the Threaded Algebraic Space-Time (TAST) codes to asynchronous transmission \cite{Damen:2007}. The delay-tolerant TAST codes are based on three different thread structures where the threads are separated by using different algebraic or transcendental numbers that guarantee a non-zero determinant of the codewords distance matrix. An extension of this TAST framework to minimum delay length codes was considered in \cite{Torbatian:2008}. 

Meanwhile, perfect space-time block codes that are optimal codes originally constructed for MIMO systems \cite{GC, PC, Elia:2006-1, Elia:2007}, were also investigated for wireless relay networks. In \cite{Yang:2007, Oggier:2008}, the authors provided optimal coding schemes in the sense of DMT tradeoff \cite{Zheng:2003}, based cyclic division algebras for any number of users and for different cooperative strategies. Nevertheless, all these schemes assumed perfect synchronization between users. Then, it was in \cite{Elia:2006-2} that Petros and Kumar discussed the delay-tolerant version of the optimal perfect code variants for asynchronous transmission. They stated that delay-tolerant diagonally-restricted CDA codes and delay-tolerant full-rate CDA codes can be obtained from previous designs by multiplying the codeword matrix by a random unitary matrix. This matrix can be taken specifically from an infinite set of unitary matrices that do not have elements in the code field.

In this paper, we construct delay-tolerant distributed codes based on the perfect codes algebras from a different point of view. The new construction is obtained from the tensor product of two number fields, one of them being the field used for the perfect code. The codes are designed in such a way to maintain the same properties of their corresponding perfect codes in the synchronous transmission, namely full-rate, full-diversity and non-vanishing minimum determinant. In addition, unlike the perfect codes, the new codes preserve the full diversity in the asynchronous transmission.

\section{Background}

Before addressing the STC construction, we dedicate this section to briefly review the remarkable properties of the perfect codes as analyzed in \cite{GC, PC, Elia:2006-1, Elia:2007}. Then, following the framework of \cite{Li:2005}, we present the cooperative communication model of interest.

\subsection{Perfect Space-Time Block Codes}\label{PerfectCodes}
The concept of Perfect Code was originally proposed in \cite{GC, PC} for $N_t=2,3,4,6$ transmit antennas to describe a square $N_t\times N_t$ linear dispersion STC $\mathcal{C}$. The perfect codes are constructed from cyclic division algebras  $\mathcal{A}(\mathbb{K}/\mathbb{F}, \sigma, \gamma)$ of degree $n=N_t$ defined by

\begin{enumerate}
  \item[-] $\mathbb{K}$ and $\mathbb{F}$ are number fields and $\mathcal{O}_\mathbb{K}, \mathcal{O}_{\mathbb{F}}$ the corresponding ring of integers. $\mathbb{F}$ is called the base field and taken as $\mathbb{F}=\mathbb{Q}(i)$ or $\mathbb{F}=\mathbb{Q}(j)$ since the ST code transmits $q$-QAM or $q$-HEX information symbols for $N_t=2,4$ or $N_t=3,6$, respectively. Thus, the constellations can be seen as finite subsets of the ring of Gaussian integers $\mathcal{O}_{\mathbb{F}}=\mathbb{Z}[i]$ or Eisenstein integers $\mathcal{O}_{\mathbb{F}}=\mathbb{Z}[j]$ $(i=\sqrt{-1}, j=e^{2\pi i/3})$, respectively.
  \item[-] $\mathbb{K}/\mathbb{F}$ is a cyclic Galois extension of $\mathbb{F}$ of degree $[\mathbb{K}:\mathbb{F}]=n$ with $\mathbb{K}=\mathbb{Q}(i,\theta)$ or $\mathbb{K}=\mathbb{Q}(j,\theta)$ a field extension appropriately chosen in order to get an existing lattice and a division algebra, and $\theta$ an algebraic number.
  \item[-] $\sigma$ is the generator of the Galois group $\mathrm{Gal}(\mathbb{K}/\mathbb{F})$, $\mathrm{Gal}(\mathbb{K}/\mathbb{F})=\langle \sigma \rangle = \{\sigma^{k}\}_{k=1}^n$.
      For an element $x \in \mathbb{K}$, the conjugates of $x$ are $\sigma^k(x)$. So, the norm $N_{\mathbb{K}/\mathbb{F}}$ and the trace $\mathrm{Tr}_{\mathbb{K}/\mathbb{F}}$ are defined respectively as
      \begin{equation}
      N_{\mathbb{K}/\mathbb{F}}(x)=\prod\limits_{k=1}^{n}\sigma^k(x), ~~~ \mathrm{Tr}_{\mathbb{K}/\mathbb{F}}(x)=\sum\limits_{k=1}^{n}\sigma^k(x).
      \end{equation}
  \item[-] $\gamma \in \mathbb{F}^{*}=\mathbb{F}/\{0\}$ the set of non-zero elements of $\mathbb{F}$. It is a non-norm element suitable for the cyclic extension $\mathbb{K}/\mathbb{F}$ \cite{PC}.

The cyclic division algebra is then expressed as a right $\mathbb{K}$-space
\begin{eqnarray}
&\mathcal{A}_1=\mathbb{K} \oplus e\mathbb{K} \oplus e^2\mathbb{K} \oplus\ldots \oplus e^{n-1}\mathbb{K}\\
&\mathrm{with}~~ e\in \mathcal{A}, e^n=\gamma \in\mathbb{K}, \gamma\neq 0 ~~\mathrm{and}~~ \lambda e=e\sigma(\lambda)~~ \mathrm{for~all}~~ \lambda \in\mathbb{K}.
\end{eqnarray}
\end{enumerate}

The Perfect Codes $\mathcal{P}$ satisfy the criteria:
\begin{enumerate}
  \item[\textbullet] \textbf{Full-rate}: The code transmits $N_t^2$ symbols drawn from QAM or HEX constellation and thus has a rate of $R=N_t$ symbols per channel use (spcu).
  \item[\textbullet] \textbf{Full-diversity}: According to the rank criterion \cite{TAROKH:1998}, the determinant of the codeword distance matrix $\mathbf{A}=(\mathbf{X}_i-\mathbf{X}_j)(\mathbf{X}_i-\mathbf{X}_j)^{\dag}$ for any two distinct codewords is non-zero. By code linearity, it can be reduced to
      \begin{equation}\label{DeterminantCondition}
      \det(\mathbf{A})\neq0 ~~\Rightarrow ~~ \det(\mathbf{X}\mathbf{X}^{\dag})=|\det(\mathbf{X})|^2\neq0,~~\mathbf{X}\neq\mathbf{0}, \mathbf{X}\in\mathcal{C}
      \end{equation}
  \item[\textbullet] \textbf{Non-vanishing minimum determinant}: The minimum determinant of any codeword distance matrix, prior to SNR normalization, is lower bounded by a constant $\psi$ that is independent of the constellation size
      \begin{equation}
      \delta(\mathcal{C})=\min\limits_{\mathbf{0}\neq\mathbf{X}\in\mathcal{C}}|\det(\mathbf{X})|^2\geq\psi>0
      \end{equation}
  \item[\textbullet] \textbf{Cubic shaping}: The QAM or HEX constellations are normalized according to the power at the transmitter so that the real vectorized codeword vectors are isomorphic to cubic lattices $\mathbb{Z}^{2N_t^2}$ or $\mathbb{A}^{2N_t^2}$. In other words, the rotation matrix $\mathbf{M}$ encoding the information symbols into each layer is required to be unitary to guarantee the energy efficiency of the codes. The shaping constraint leads thus to two other properties. The first one is the \textbf{Uniform average transmitted energy per antenna}. The second one is the \textbf{Information losslessness} as the unitary linear dispersion matrix $\mathbf{M}$ allows to preserve the mutual information of the MIMO channel.
\end{enumerate}

Thanks to prominent results on diversity-multiplexing tradeoff \cite{Zheng:2003}, the perfect codes also verify two other equivalent properties:
\begin{enumerate}
  \item[\textbullet] \textbf{DMT optimality}: In \cite{Elia:2006-1}, Elia \etal~ proved that the full-rate STCs from cyclic division algebra having NVD property achieve the optimal DMT in Rayleigh fading channel.
  \item[\textbullet] \textbf{Approximate universality}: Being CDA-based codes with NVD property, the perfect codes are approximately universal and achieve DMT for arbitrary channel fading distribution.
\end{enumerate}

Satisfying all these criteria, the perfect codes showed to improve the performance in terms of error probability upon the best known codes.

\subsection{Cooperative System Model}\label{CoopSysModel}

\begin{figure}[h] %
\centering %
\includegraphics[width = 0.8\textwidth]{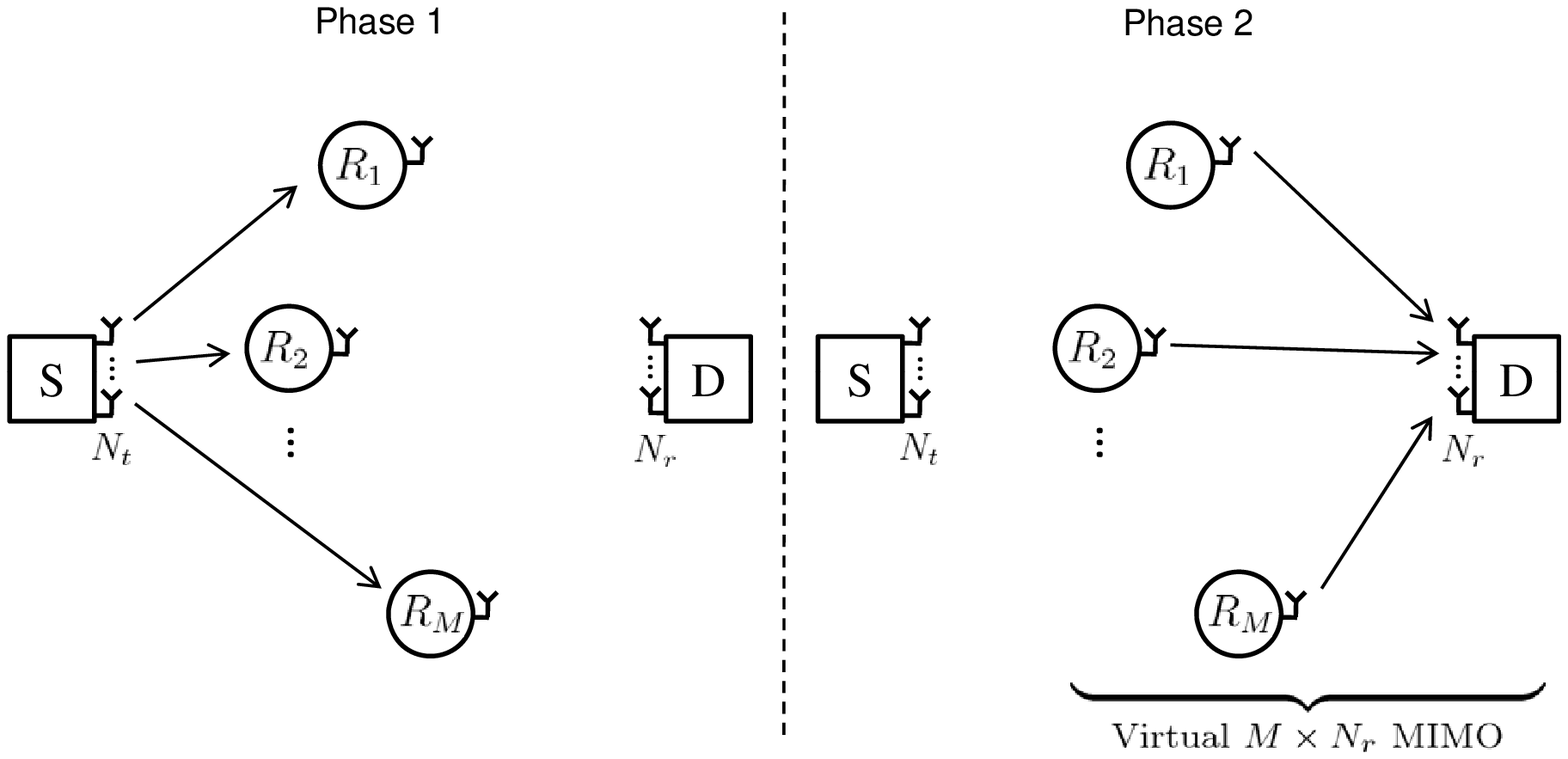}\vspace{-8pt} %
\caption{Cooperative System Architecture} %
\label{CoopSyst}%
\end{figure}

In the sequel, we consider a cooperative system with a source $\mathrm{S}$ communicating to a destination $\mathrm{D}$ via $M$ relays $R_i$ in two phases as in Figure \ref{CoopSyst}, and without direct links between the source and the destination. In the first phase, the source broadcasts its message to the potential relays. In the second phase, the relays use the DF protocol to detect the source message then if successfully detected transmit it to the destination. We assume that all the $M$ relays are able to achieve error free decoding which could be possible by selecting the source-relays links, and consider only the links that are not in outage. Note that it could also be possible that not all the relays may successfully decode the original message, so the number of transmitting relays is usually assumed as a random variable. Since the relays transmission overlap in time and frequency, they can cooperatively implement a distributed space-time code.

Considering only the second phase of transmission, the system is equivalent to a MIMO scheme where the distributed $M\times T$ perfect space-time code $\mathcal{P}$ is used by the relays, with $M$ transmit antennas one by relay, and $N_r$ receive antennas at the destination. Every time slot $t, t=1\ldots T$, the relays send the $M\times 1$ $t^{th}$ column vector $\mathbf{X}_t$ of the codeword $\mathbf{X}$ and the destination receives
\begin{equation}\label{CoopModel}
\mathbf{Y}_t=\mathbf{H}_t\mathbf{X}_t+\mathbf{W}_t,~~~\mathbf{Y}_t,\mathbf{W}_t\in\mathbb{C}^{N_r\times 1}
\end{equation}
where $\mathbf{W}_t$ is the additive white Gaussian noise with i.i.d complex Gaussian variables with zero-mean and variance $N_0$, $\sim \mathcal{N}_c(0,N_0)$, $N_0=2\sigma^2$, $\sigma^2$ being the noise variance per real dimension. $\mathbf{H}_t$ represents the $N_r\times M$ complex channel matrix modeled as i.i.d Gaussian random variables with zero mean and unit variance $\sim \mathcal{N}_c(0,1)$. The channel is assumed quasi-static with constant fadings during a transmitted codeword and independent fadings between subsequent codewords. Dealing with square STCs $(M=T)$, the codeword matrix $\mathbf{X}_t$ contains $M^2$ information symbols $s_1, \ldots, s_{M^2}$ carved from two-dimensional QAM or HEX finite constellations denoted by $\mathcal{S}$.

\subsection{Asynchronous Cooperative Diversity}
The above expression \eqref{CoopModel} is valid only when relays are synchronized. In the presence of asynchronicity, the codeword transmission is spanned on more than $T$ symbol intervals due to delays. Although the symbol synchronization is not required, we assume that the relays are synchronized at the frame-codeword level, which can be provided by means of network feedback signaling from the destination. Therefore, the start and the end of each codeword are aligned for different relays by transmitting zero symbols, and hence there is no interference between codewords transmission. We further assume that the timing errors between different relays are integer multiples of the symbol duration and the fractional timing errors are absorbed in the channel dispersion. In the codeword matrix, these delays are also filled with zeros; they are known at the receiver but not at the transmitting relays \cite{Li:2005}.

Denoting a delay profile by $\bm{\mathfrak{d}}=(\mathfrak{d}_1,\mathfrak{d}_2, \ldots, \mathfrak{d}_M)$, a delay $\mathfrak{d}_i$ corresponds to the relative delay of the received signal from the $i^{th}$ relay as referenced to the earliest received relay signal. Let $\mathfrak{d}_{\max}$ denotes the maximum of the relative delays, then from the receiver perspective, the $M\times (T+\mathfrak{d}_{\max})$ codeword matrix was sent instead of the $M\times T$ space-time code.

\subsection{Motivation of the Code Construction}

The diversity order of any space-time code is defined by the minimum rank of the distance codeword matrix over all pairs of distinct codewords \cite{TAROKH:1998}. The distributed $M\times T$ perfect codes $\mathcal{P}$ are full-rate full-diversity for the synchronous transmission between the relays and the destination. Note that in general, a transmission between source, half-duplex relays and destination will result in rate loss. When asynchronicity is introduced, the code is no more full-rate since it is spanned on $(T+\mathfrak{d}_{\max})$ time instants. Moreover, certain delay profiles $\bm{\mathfrak{d}}$ can result in linearly dependent rows, thus the code will loose its full-diversity property. Let us illustrate this by the following example.

\paragraph*{Example of Golden Code}
We consider the distributed $2\times 2$ Golden code transmitting $4$ information QAM symbols $s_1,s_2,s_3,s_4$ from two synchronized relays with the codeword matrix.
\begin{equation}\label{GoldenMatrix}
\mathbf{X}_s = \frac{1}{\sqrt{5}}\left[
\begin{array}{cc}
  \alpha(s_1+s_2\theta) & \alpha(s_3+s_4\theta) \\
  i\bar{\alpha}(s_3+s_4\bar{\theta}) & \bar{\alpha}(s_1+s_2\bar{\theta})
\end{array}
\right]
\end{equation}
The Golden code is designed on a cyclic field extension of degree $2$ over the base field $\mathbb{Q}(i)$. Using the generator matrix of the corresponding complex $2$-dimensional lattice, the codeword elements are lattice points obtained by  linear combination of pairs of symbols.

Now, let the first relay be delayed by one symbol period with respect to the second $\mathbf{\mathfrak{d}}=(1,0)$, such that the new asynchronous codeword matrix be
\begin{equation}
\mathbf{X}_a = \frac{1}{\sqrt{5}}\left[
\begin{array}{ccc}
  0 & \alpha(s_1+s_2\theta) & \alpha(s_3+s_4\theta) \\
  i\bar{\alpha}(s_3+s_4\bar{\theta}) & \bar{\alpha}(s_1+s_2\bar{\theta}) & 0
\end{array}
\right]
\end{equation}

Suppose we have two distinct codewords $\mathbf{X}_1$ and $\mathbf{X}_2$ with $s_{1,1}=-s_{1,2}=-s_1$ and the other symbols equal \ie $s_{i,1}=s_{i,2}, i=2\ldots4$. The difference between matrix codewords is defined in both synchronous and asynchronous cases as
\begin{equation}
\Delta(\mathbf{s})_s=\left[
                       \begin{array}{cc}
                         2\alpha s_1 & 0 \\
                         0 & 2\bar{\alpha}s_1 \\
                       \end{array}
                     \right]~,~~~
\Delta(\mathbf{s})_a=\left[
                       \begin{array}{ccc}
                         0 & 2\alpha s_1 & 0 \\
                         0 & 2\bar{\alpha}s_1 & 0 \\
                       \end{array}
                     \right]
\end{equation}
It can be seen that $\Delta(\mathbf{s})_s$ is a full-rank matrix whereas $\Delta(\mathbf{s})_a$ has rank one, so the Golden code is not a delay-tolerant code.

In fact, it can be seen from the asynchronous codeword matrix $\mathbf{X}_a$ that some symbols are aligned at the same instant due to delays loosing thus diversity. In order to resolve this problem of rank deficiency, our solution consists in transmitting from each antenna (relay) at each transmission time a different combination of all the $4$ information symbols. This way, in the presence of delays, we ensure that any combined symbol sent from the $2$ relays arrives at the destination in at least $2$ different instants, hence guaranteeing the full-diversity order of the space-time code.

A new $2\times2$ STC will have then the shifted codeword matrix
\begin{equation}
\mathbf{X}_a = \left[
\begin{array}{ccc}
  0 & f_1(s_1,s_2,s_3,s_4) & f_2(s_1,s_2,s_3,s_4) \\
  f_3(s_1,s_2,s_3,s_4) & f_4(s_1,s_2,s_3,s_4) & 0
\end{array}
\right].
\end{equation}
Now, to get these $4$ linear combinations of the $4$ symbols, we need a higher dimensional lattice $(n=4)$ compared to the $2$-dimensional lattice used for the Golden code. So, we propose to obtain the corresponding $4\times 4$ lattice generator matrix by the tensor product of two field extensions of $\mathbb{Q}(i)$, one of them being the field extension of the Golden code.

Following this idea, we aim at constructing, in general, new $M\times M$ codes that are based on CDA of the $M\times M$ perfect codes such that they maintain the same optimal properties as perfect codes in the synchronous case. But also, these codes preserve their full-diversity in asynchronous transmission and thus are delay-tolerant for arbitrary delay profile.

\section{Construction of Delay-Tolerant Distributed Codes \\based Perfect Codes Algebras}

\subsection{General Construction}\label{sec732}
The approach consists in constructing a division algebra isomorphic to the tensor product (also called Kronecker product or cross-product) of two number fields of lower degrees. Other constructions based on the crossed-product algebras have been investigated in \cite{Shashidhar:2006, Berhuy:2007} either for prime or coprime degrees of the composite algebras. In these constructions, the space-time code was built on the cyclic product algebra. However, in the present construction, the higher degree algebra is only used to derive appropriately the space-time code.

Since we intend to construct a full-rate $M\times M$ space-time code that is based on the CDA of the full-rate $M\times M$ perfect code, then the first algebra to be considered is the cyclic division algebra of the perfect code $\mathcal{A}_1(\mathbb{K}_1/\mathbb{F},\sigma_1,\gamma_1)$ of degree $n_1=M$ over the base field $\mathbb{F}$.
For sake of simplicity, we analyze in the sequel the case of Gaussian Field $\mathbb{F}=\mathbb{Q}(i)$ to explain the construction. Indeed, we consider the cyclic field extension $\mathbb{K}_1=\mathbb{Q}(i,\theta_1)$ of degree $n_1=M$ over $\mathbb{F}$, $\theta_1$ being an algebraic number. The principal ideal $\mathcal{I}_{\mathbb{K}_1}\subseteq \mathcal{O}_{\mathbb{K}_1}$ is generated by an element $\alpha$ and its integral basis is $\mathrm{B}_1=(v_1, v_2, \ldots, v_{M})$ (or if unitary, it is given by $\mathrm{B}_1=(\alpha,\alpha\theta_1,\ldots,\alpha\theta_1^{M-1})$). 
The basis of the complex algebraic lattice $\Lambda(\mathcal{I}_{\mathbb{K}_{1}})$ is obtained by applying the canonical embedding to $\mathrm{B}_{1}$.
Consequently, the generator matrix corresponds to the rotation matrix in $\mathbb{Z}[i]^M$
\begin{equation}
\mathbf{M}_1=\frac{1}{\sqrt{p_1}}\left[
  \begin{array}{cccc}
    v_1 & v_2 & \ldots & v_{M} \\
    \sigma_1(v_1) & \sigma_1(v_2) & \ldots & \sigma_1(v_{M}) \\
    \vdots & \vdots & \ddots & \vdots \\
    \sigma_1^{M-1}(v_1) & \sigma_1^{M-1}(v_2) & \ldots & \sigma_1^{M-1}(v_{M}) \\
  \end{array}
\right],
\end{equation}
where $\sqrt{p_1}$ is a normalization factor used to guarantee the matrix unitarity.

Now, we consider another Galois extension $\mathbb{K}_{2}$ over $\mathbb{F}$ of the same degree $n_2=M$ such that its discriminant is coprime to the one of $\mathbb{K}_{1}$ i.e., $(d_{\mathbb{K}_{1}}, d_{\mathbb{K}_{2}})=1$. Let $\mathbb{K}_{2}=\mathbb{Q}(\theta_2)$ with $\theta_2$ an algebraic number. The Galois group is generated by $\sigma_{2}$ as $\mathrm{Gal}(\mathbb{K}_{2}/\mathbb{F})=\langle\sigma_2\rangle$. The principal ideal of the algebra is such that $\mathcal{I}_{\mathbb{K}_2}=\mathcal{O}_{{\mathbb{K}}_2}$ and thus its integral basis is given by $\mathrm{B_2}=(1, \theta_2, \ldots, \theta_2^{M-1})$. The canonical embedding of $\mathrm{B_2}$ gives another complex rotated lattice of $\mathbb{Z}[i]^M$ that is generated by the unitary matrix $\mathbf{M}_2$ with $\sqrt{p_2}$ the normalization factor,

\begin{equation}
\mathbf{M}_2=\frac{1}{\sqrt{p_2}}\left[
  \begin{array}{cccc}
    1 & \theta_2 & \ldots & \theta_2^{M-1} \\
    1 & \sigma_2(\theta_2) & \ldots & \sigma_2(\theta_2^{M-1}) \\
    \vdots & \vdots & \ddots & \vdots \\
    1 & \sigma_2^{M-1}(\theta_2) & \ldots & \sigma_2^{M-1}(\theta_2^{M-1}) \\
  \end{array}
\right].
\end{equation}

\begin{figure}[h] %
\centering{ %
\includegraphics[width = 0.5\textwidth]{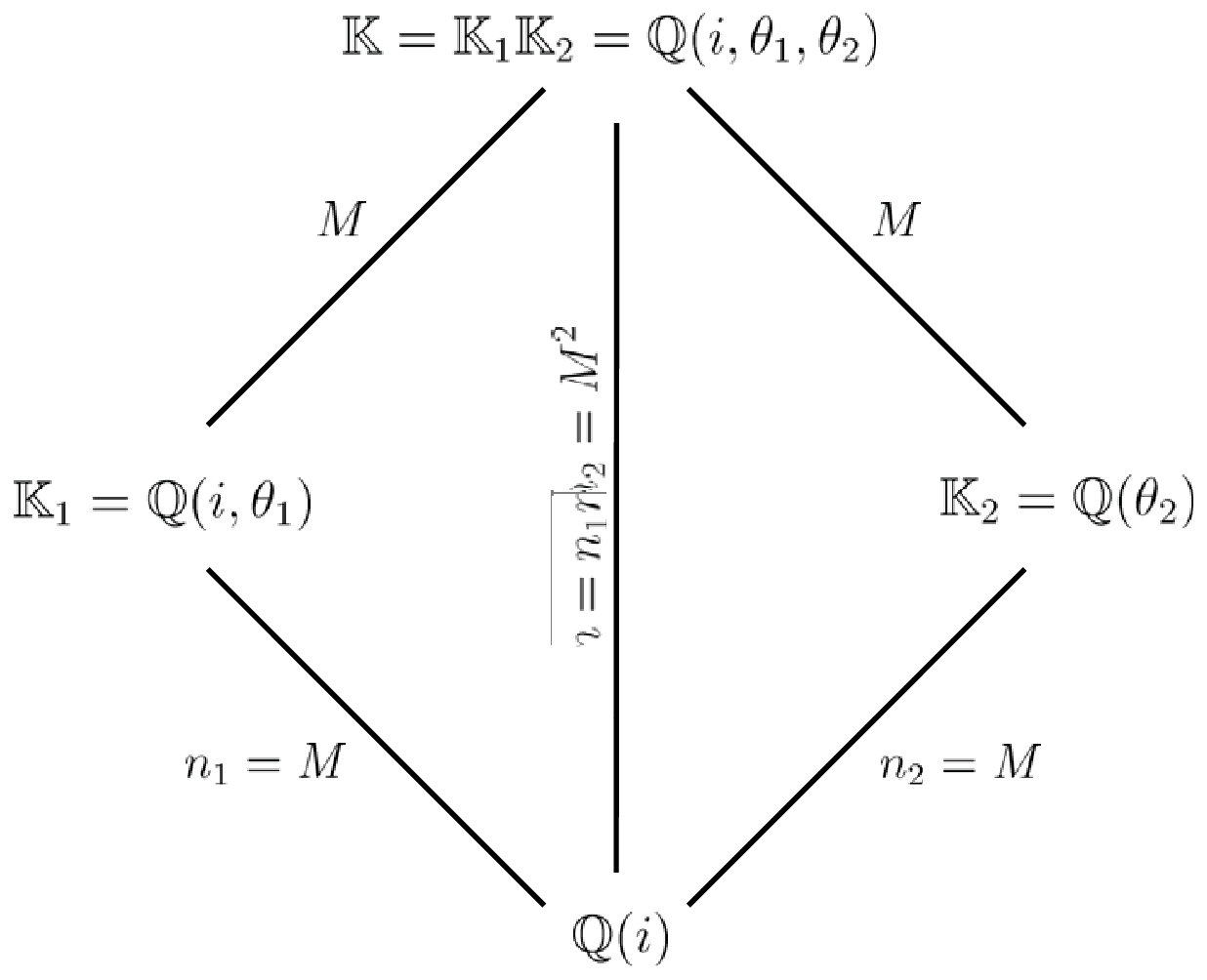} \vspace{-8pt}%
\caption{Compositum Field of the Tensor Product Algebra} %
\label{TensorProduct}}%
\end{figure}

The tensor product of both field extensions allows to build a rotated lattice in higher dimension corresponding to the complex $M^2\times M^2$ unitary matrix $\mathbf{M}$ based on the previous $M \times M$ constructions. According to \cite{Bayer:2004},
\paragraph*{Proposition $1$}: Let $\mathbb{K}$ be the compositum of the above Galois extensions, $\mathbb{K}=\mathbb{K}_{1}\mathbb{K}_{2}=\mathbb{Q}(i, \theta_1, \theta_2)$ of order $n=n_1n_2=M^2$ over $\mathbb{F}$ as presented in Figure \ref{TensorProduct}.

Since $\mathbb{K}_{1}$ and $\mathbb{K}_{2}$ have coprime discriminants, the corresponding lattice generator matrix can be obtained as the tensor product of the previous unitary generator matrices.
\begin{equation}
\mathbf{M}=\mathbf{M}_2\otimes\mathbf{M}_1=\frac{1}{\sqrt{p_1p_2}}\cdot \nonumber
\end{equation}
\begin{equation}
\left[\!
\renewcommand{\arraystretch}{0.6}
       \begin{array}{ccccccc}
         v_1 & \cdots & v_{M} & \cdots & v_1\theta_2^{M-1} & \cdots & v_{M}\theta_2^{M-1}\\
         \vdots & \ddots & \vdots & \ddots & \vdots & \ddots & \vdots\\
         \sigma_1^{M-1}(v_1) & \cdots & \sigma_1^{M-1}(v_{M}) & \cdots & \sigma_1^{M-1}(v_1)\theta_2^{M-1} & \cdots & \sigma_1^{M-1}(v_{M})\theta_2^{M-1}\\
         \vdots & \ddots & \vdots & \ddots & \vdots & \ddots &  \vdots\\
         v_1 & \cdots & v_{M} & \cdots & v_1\sigma_2^{M-1}(\theta_2^{M-1}) & \cdots & v_{M}\sigma_2^{M-1}(\theta_2^{M-1})\\
         \vdots & \ddots & \vdots & \ddots & \vdots & \ddots & \vdots\\
         \sigma_1^{M-1}(v_1) & \cdots & \sigma_2^{M-1}(v_{M}) & \cdots & \sigma_1^{M-1}(v_1)\sigma_2^{M-1}(\theta_2^{M-1}) & \cdots & \sigma_1^{M-1}(v_{M})\sigma_2^{M-1}(\theta_2^{M-1})\\
        \end{array}
\!\right]\vspace{-10pt}
\end{equation}
Consequently,
\paragraph*{Proposition $2$}: Let $m_j=[\mathbb{K}:\mathbb{K}_j]=n/n_j,~j=1,2$ the order of the extensions, then the discriminant of $\mathbb{K}$ is $d_{\mathbb{K}}=d_{\mathbb{K}_1}^{m_1}d_{\mathbb{K}_2}^{m_2}$. The minimum product distance of the lattice is derived from the discriminant of $\mathbb{K}$ as
\begin{equation}\label{equ76}
d_{\mathrm{p,min}}=\frac{1}{\sqrt{d_{\mathbb{K}}}}=\frac{1}{\sqrt{d_{\mathbb{K}_1}^{m_1}d_{\mathbb{K}_2}^{m_2}}}
\end{equation}

Using the matrix $\mathbf{M}$, the space-time coded components are given by the linear combination $\mathbf{x}=\mathbf{M}\mathbf{s}$ where $\mathbf{s}=[s_1,s_2, \ldots, s_{M^2}]^T$ is the information symbol vector carved from a $q$-QAM$^{M^2}$ constellation $(\in \mathbb{Z}[i]^{M^2})$. Then, the space-time codeword matrix is defined by distributing the components with appropriate constant factors $\phi_{l}, l=1,\ldots,M^2$. It can be represented as a Hadamard product
\begin{eqnarray}
\mathbf{X} & = &[{\Phi}]\bullet[\mathbf{x}]=\left[
             \begin{array}{cccc}
               \phi_1x_1 & \phi_{M+1}x_{M+1} & \cdots & \phi_{M(M-1)+1}x_{M(M-1)+1} \\
               \phi_2x_2 & \phi_{M+2}x_{M+2} & \cdots & \phi_{M(M-1)+2}x_{M(M-1)+2} \\
               \vdots & \vdots & \ddots & \vdots \\
               \phi_Mx_M & \phi_{2M}x_{2M} & \cdots & \phi_{M^2}x_{M^2} \\
             \end{array}
           \right]
\end{eqnarray}

The key idea in the code construction is to determine the coefficients $\phi_l$ that allow one to preserve the same properties of the corresponding perfect codes in synchronous transmission (Section \ref{PerfectCodes}).
\begin{enumerate}
\item[\textbullet] On one side, it can be seen that the new code transmits $M^2$ information symbols and thus is \textit{full-rate} with $R=M$ spcu for a relays-destination transmission phase.
\item[\textbullet] On the other side, we need to find the $\phi_l$ factors that satisfy the rank criterion (\ref{DeterminantCondition}) in order to have \textit{full-diversity} codes.
\item[\textbullet] Moreover, the perfect codes have \textit{non-vanishing minimum determinants}. Then, we are interested in deriving $M\times M$ ST codes that have not only non-zero determinants, but also these determinants do not vanish when constellation size increases.
\item[\textbullet] In order to guarantee \textit{uniform energy distribution} in the codeword, we ask that $\phi_l$ verify $|\phi_l|=1$. Choosing further the coefficients $\phi_l \in \mathcal{O}_{\mathbb{F}}=\mathbb{Z}[i]$ yields better determinants as obtained for the non-norm elements $\gamma$ of the perfect codes \cite{PC}. This restricts the values of $\phi_l$ to $\phi_l=\pm 1,\pm i$.
\item[\textbullet] It can also be noticed that the new code satisfies the \textit{cubic shaping property} since the generator matrix $\mathbf{M}$ of the $M^2$-dimensional lattice is unitary, and hence the code is \textit{information lossless}.
\end{enumerate}
In addition, when asynchronicity between relays is involved, the rank criterion should be also verified for the shifted matrix and another criterion will be analyzed that is the non-zero product distance of the codeword matrix in order to prove that the new codes are delay-tolerant, and thus keep their full-diversity in asynchronous transmission.

\section{New Delay-Tolerant Codes from $2,3,4$-dimensional Perfect Codes}
Based on the previous approach, we consider the perfect codes proposed in \cite{GC, PC} for dimensions $M=2,3,4$ to construct the new delay-tolerant codes. Then, in the next section, we apply this construction for the perfect codes presented for any number of antennas in \cite{Elia2:2005}.

\subsection{$2\times2$ Code based on Golden Code}
The Golden Code was constructed in \cite{GC} using the cyclic division algebra $\mathcal{A}_{1}(\mathbb{K}_{1}/\mathbb{Q}(i),\sigma_{1},i)$ of degree $2$ over $\mathbb{F}=\mathbb{Q}(i)$. $\mathbb{K}_{1}=\mathbb{Q}(i,\sqrt{5})$ is a Galois extension of degree $2$. It is a $2$-dimensional vector space of $\mathbb{Q}(i)$ with basis $\mathrm{B}=(1,\theta_1)$, $\theta_1=\frac{1+\sqrt{5}}{2}$ being the Golden number. Its Galois group $\mathrm{{Gal}}({\mathbb{K}_{1}}/{\mathbb{F}})$ is generated by $\sigma_{1}: \sqrt{5}\mapsto-\sqrt{5}$. In order to get a rotated lattice $\Lambda(\mathcal{I}_{\mathbb{K}_{1}})$ of $\mathbf{Z}[i]^2$, the principal ideal $\mathcal{I}_{\mathbb{K}_{1}}=\alpha\mathcal{O}_{\mathbb{K}_1}$ generated by $\alpha=1+i-i\theta_1$ was found. Its basis is $\mathrm{B_1}=(\alpha, \alpha\theta_1)$ and its unitary generator matrix is given by
\begin{equation}
\mathbf{M}_1=\frac{1}{\sqrt{5}}\left[
        \begin{array}{cc}
          \alpha & \alpha \theta_{1}\\
         \bar{ \alpha} & \bar{\alpha}\bar{\theta_{1}} \\
        \end{array}
      \right],
\end{equation}
with $\bar{\theta_{1}}=\frac{1-\sqrt{5}}{2}$ and $\bar{\alpha}=1+i-i\bar{\theta_{1}}$ the respective conjugates of $\theta_1$ and $\alpha$.

Let $\mathbb{K}_{2}=\mathbb{Q}(\theta_2)$ the cyclotomic extension of degree $2$ over $\mathbb{F}$ with $\theta_2=\zeta_8=e^{{i\pi}/{4}}$ the primitive $8^{th}$ root of unity. Its discriminant $d_{\mathbb{K}_{2}}=4$ and it is coprime to the one of $\mathbb{K}_{1}$ since $d_{\mathbb{K}_{1}}=5$. The Galois group $\mathrm{{Gal}}({\mathbb{K}_{2}}/{\mathbb{F}})$ is generated by $\sigma_{2}: \zeta_8\mapsto-\zeta_8$ and the integral basis of $\mathbb{K}_2$ is $\mathrm{B_2}=(1,\zeta_8)$. The corresponding unitary generator matrix is
 \begin{equation}
\mathbf{M}_2=\frac{1}{\sqrt{2}}\left[
        \begin{array}{cc}
          1 & \zeta_8\\
         1  & -\zeta_8 \\
        \end{array}
      \right]
\end{equation}

Therefore, $\mathbb{K}=\mathbb{K}_1\mathbb{K}_2=\mathbb{Q}(i, \theta_1,\zeta_8)$ is the compositum of Galois extensions of degree $2$ each, with coprime discriminants. The $4 \times 4$ unitary matrix is obtained by the tensor product of previous matrices as
\begin{equation}
\mathbf{M}_4=\frac{1}{\sqrt{10}}\left[
   \begin{array}{cccc}
    \alpha & \alpha\theta_1 & \alpha\zeta_8 & \alpha\theta_1\zeta_8 \\
     \bar{\alpha} & \bar{\alpha}\bar{\theta_1} & \bar{\alpha}\zeta_8 & \bar{\alpha}\bar{\theta_1}\zeta_8 \\
      \alpha & \alpha\theta_1 & -\alpha\zeta_8 & -\alpha\theta_1\zeta_8 \\
      \bar{\alpha} & \bar{\alpha}\bar{\theta_1} & -\bar{\alpha}\zeta_8 & -\bar{\alpha}\bar{\theta_1}\zeta_8 \\
   \end{array}
 \right]
\end{equation}
and the codeword matrix is defined by
\begin{equation}\label{equ713}
\bm{\Gamma}(\mathbf{s}) = \left[\begin{array}{cc}
  \phi_1x_1 &  \phi_3x_3 \\
  \phi_2 x_2 &  \phi_4x_4
\end{array}\right],
\end{equation}
where $x_i$ are the components of the vector $\mathbf{M}_4\mathbf{s}$ with $s_1,s_2,s_3,s_4$ are $q$-QAM symbols. We propose now to determine the coefficients $\phi_l, l=1,\ldots,4$ that satisfy the non-vanishing determinant criterion.\smallskip

\subsubsection{Non-vanishing minimum determinant}

The determinant of this codeword matrix equals
\begin{equation}
\delta(\mathbf{s}) = \phi_1\phi_4 x_1 x_4 -\phi_2\phi_3 x_2 x_3.
\end{equation}
By developing $x_1 x_4$ and $x_2 x_3$, we obtain
\begin{eqnarray}
x_1 x_4 &= &\frac{1}{10}N_{\mathbb{K}_1/\mathbb{F}}(\alpha)\left( G(s) + \frac{1+i}{\sqrt{2}}\sqrt{5}s_1s_4 -\frac{1+i}{\sqrt{2}}\sqrt{5}s_2s_3\right),\\
x_2 x_3 &= &\frac{1}{10}N_{\mathbb{K}_1/\mathbb{F}}(\alpha)\left( G(s) - \frac{1+i}{\sqrt{2}}\sqrt{5}s_1s_4+\frac{1+i}{\sqrt{2}}\sqrt{5}s_2s_3\right),
\end{eqnarray}
with
\begin{equation}
N_{\mathbb{K}_1/\mathbb{F}}(\alpha)=\alpha\bar{\alpha}=2+i~~\mathrm{and}~~
G(s)= s_1^2 - s_2^2 -is_3^2 + is_4^2 +s_1s_2 -is_3s_4.
\end{equation}
It is interesting to note that the Golden codeword given by matrix (\ref{GoldenMatrix}) has a determinant of
\begin{equation}
\delta'(\mathbf{s}) = \frac{1}{5}N_{\mathbb{K}_1/\mathbb{F}}(\alpha)G(s).
\end{equation}

Therefore, by choosing $\phi_1=\phi_3=\phi_4=1$ and $\phi_2=-1$, the determinant of the new code is equal to the Golden code determinant, and does not vanish when increasing the size of the QAM constellation carved from $\mathbb{Z}[i]$. Hence, the new code achieves the diversity-multiplexing tradeoff \cite{Zheng:2003, Elia:2006-1}.

It can also be noticed that the coefficients $\phi_l$ can be changed equivalently to the coefficients of the Fourier matrix $F_{n}=(w^{jk})$ where $w=e^{2i \pi/n}$ is the primitive $n^{th}$ root of unity. For dimension $2$, we have
\begin{equation}
[{\Phi}]= \left[
            \begin{array}{cc}
              1 & 1 \\
              1 & -1 \\
            \end{array}
          \right]
\end{equation}

Furthermore, we have find fixed unitary matrices $\mathbf{U}$ and $\mathbf{V}$ such that $\bm{\Gamma}=\mathbf{U}\mathbf{G}\mathbf{V}$ for all values of $s_1, s_2, s_3, s_4$ with
\begin{equation}
\mathbf{U}=\left[\begin{array}{cc}
  \zeta_8 & 0 \\
  0 & -1
\end{array}\right]\,,\,
\mathbf{V}=\frac{1}{\sqrt{2}}\left[\begin{array}{cc}
 -i\zeta_8 & -i\zeta_8 \\
  1 & -1
\end{array}\right].
\end{equation}\smallskip

\subsubsection{Delay-tolerance}

In the distributed setup, each row of the code matrix is transmitted by a different relay (Section \ref{CoopSysModel}). In practical scenarios, the two relays do not share a common timing reference, and therefore, the arrival of packets is not synchronous. As we assume synchronization at the symbol level, the distributed code can still achieve full diversity if the differences between matrix codewords are full rank even when the different rows are arbitrarily shifted. In what follows, we prove that the new code $\bm{\Gamma}$ satisfy this condition.

Consider the shifted codeword matrix of $\bm{\Gamma}$
\begin{equation}
\bm{\Gamma}_a=\left[ \begin{array}{ccc}
            0 & x_1 & x_3 \\
            -x_2 & x_4 &0
        \end{array} \right],
\end{equation}
we need to guarantee that it is full rank when $\mathbf{s}\neq \mathbf{0}$ \ie for any $\mathbf{s}_1\neq \mathbf{s}_2$ from the constellation $(\mathrm{rank}=\min(M,T+1)=2)$. This restricts to show that the $2\times 2$ submatrix
\[\left[ \begin{array}{cc} 0 & x_3 \\
                                     -x_2 & 0
                                  \end{array} \right]
\]
is full rank \ie its determinant $x_2 x_3 \neq 0$ when $\mathbf{s}\neq \mathbf{0}$.

More generally, having delay profiles $\bm{\mathfrak{d}}=(1,0)$ or $(0,1)$, the problem turns to prove that the product distance in the rotated constellation associated with the matrix $\mathbf{M}_4$ of $\bm{\Gamma}$ is non-zero over $\mathbb{Z}[i]$, so that any component product is non-zero. This product distance is evaluated as
\begin{eqnarray}
d_p = \prod\limits_{j=0}^4|x_j|=|x_1x_2x_3x_4| = \Big|\frac{1}{10}\alpha\bar{\alpha}\big( G(\mathbf{s})+ \frac{1+i}{\sqrt{2}}\sqrt{5}s_1s_4 - \frac{1+i}{\sqrt{2}}\sqrt{5}s_2s_3\big)\Big|\times \nonumber \\
\Big|\frac{1}{10}\alpha\bar{\alpha}\big( G(\mathbf{s}) - \frac{1+i}{\sqrt{2}}\sqrt{5}s_1s_4 + \frac{1+i}{\sqrt{2}}\sqrt{5}s_2s_3\big)\Big| = \frac{1}{20}\left|G(s)^2-\left(\frac{1+i}{\sqrt{2}}\sqrt{5}\left(s_1s_4-s_2s_3\right)\right)^2\right|
\end{eqnarray}
with $G(\mathbf{s})= s_1^2 - s_2^2 -is_3^2 + is_4^2 +s_1s_2 -is_3s_4\in \mathbb{Z}[i]$ for $\mathbf{s} \in \mathbb{Z}[i]^4$.

As a direct consequence from the tensor product construction, Equation (\ref{equ76}) gives
\[
d_{\mathrm{p,min}}=\frac{1}{\sqrt{d_{\mathbb{K}}}}=\frac{1}{\sqrt{{5}^{2}4^{2}}}=\frac{1}{20}
\]
Thus, the minimum product distance is non-zero. It can also be verified in $d_{\mathrm{p}}$ by setting $s_1=1, s_2=s_3=s_4=0$. So, $d_p$ is non-zero unless $s_1=s_2=s_3=s_4=0$, and consequently the submatrix is full rank since $x_2 x_3 \neq 0$ unless $\mathbf{s} = \mathbf{0}$.

Therefore, the new code unlike the Golden code keeps its full-diversity in the case of asynchronous relays. However, we cannot guarantee the non-vanishing determinant property in the asynchronous case because the expression of $x_2x_3$ can be interpreted as a Diophantine approximation of $\frac{1}{\sqrt{2}}$ by rational numbers which can be made tighter by using larger constellation size.

\subsection{$3\times3$ Code based on $3\times3$ Perfect Code}
In order to construct the delay-tolerant $3\times 3$ code, we consider the base field $\mathbb{F}=\mathbb{Q}(j)$ and we use $q$-HEX symbols. Let $\theta_1=\zeta_7+\zeta_7^{-1}=2\cos(\frac{2\pi}{7})$, with $\zeta_7$ the $7^{th}$ root of unity. The $ 3\times 3$ perfect code was constructed using the cyclic division algebra $\mathcal{A}_1(\mathbb{K}_1/\mathbb{F},\sigma_1,j)$ of order $3$ \cite{PC}, where the relative extension $\mathbb{K}_1=\mathbb{Q}(j,\theta_1)$ and $\sigma_1$ the generator of the cyclic extension $\mathbb{K}_1/\mathbb{F}$ with $\sigma_1:\zeta_7+\zeta_7^{-1}\mapsto\zeta_7^2+\zeta_7^{-2}$. The integral basis is given by
$
\mathrm{B}_1=\{v_k\}_{k=1}^3=\{(1+j)+\theta_1,(-1-2j)+j\theta_1^2,(-1-2j)+(1+j)\theta_1+(1+j)\theta_1^2\}
$
 and the complex lattice $\Lambda({\mathcal{I}_{\mathbb{K}_1}})$ is a rotated version of $\mathbb{Z}[j]^3$. It is generated by
\begin{equation}
\mathbf{M}_1=\frac{1}{\sqrt{7}}\left[
               \begin{array}{ccc}
                 v_1 & v_2 & v_3 \\
                 \sigma_1(v_1) & \sigma_1(v_2) & \sigma_1(v_3) \\
                 \sigma_1^2(v_1) & \sigma_1^2(v_2) & \sigma_1^2(v_3) \\
               \end{array}
             \right].
\end{equation}

The relative discriminant of $\mathbb{K}_1$ is $d_{\mathbb{K}_1}=49$. Another extension of $\mathbb{F}$ of degree $3$ that has coprime discriminant with $\mathbb{K}_1$ is the cyclotomic extension $\mathbb{K}_2=\mathbb{Q}(\zeta_9)$ with $\zeta_9=e^{2i\pi/9}$ the primitive $9^{th}$ root of unity and $d_{\mathbb{K}_2}=27$. Its Galois group $\mathrm{Gal}(\mathbb{K}_2/\mathbb{F})$ is generated by $\sigma_2:\zeta_9\mapsto j\zeta_9$.
The integral basis of $\mathbb{K}_2$ is $\mathrm{B}_2=(1,\zeta_9,\zeta_9^2)$ and the lattice generator matrix is
\begin{equation}
\mathbf{M}_2=\frac{1}{\sqrt{3}}\left[
                                 \begin{array}{ccc}
                                   1 & \zeta_9 & \zeta_9^2 \\
                                   1 & j\zeta_9 & j^2\zeta_9^2 \\
                                   1 & j^2\zeta_9 & j\zeta_9^2 \\
                                 \end{array}
                               \right].
\end{equation}

The compositum of both extensions $\mathbb{K}=\mathbb{K}_1\mathbb{K}_2=\mathbb{Q}(j, 2\cos(\frac{2\pi}{7}),\zeta_9)$ is of order $9$ over $\mathbb{Q}(j)$. Then, the corresponding $9$-dimensional complex lattice is generated by the $9 \times 9$ unitary matrix
\begin{eqnarray}
\mathbf{M}_9 =
\frac{1}{\sqrt{21}}\left[
                     \begin{array}{ccc}
                       1 & \zeta_9 & \zeta_9^2 \\
                       1 & j\zeta_9 & j^2\zeta_9^2 \\
                       1 & j^2\zeta_9 & j\zeta_9^2 \\
                     \end{array}
                    \right]
\otimes \left[
          \begin{array}{ccc}
                         v_1 & v_2 & v_3 \\
                         \sigma_1(v_1) & \sigma_1(v_2) & \sigma_1(v_3) \\
                         \sigma_1^2(v_1) & \sigma_1^2(v_2) & \sigma_1^2(v_3) \\
          \end{array}
        \right],
\end{eqnarray}
and the $3\times 3$ space-time code is defined by the matrix
\begin{equation}
\bm{\Gamma}(\mathbf{s})=\left[
                          \begin{array}{ccc}
                            \phi_1x_1 & \phi_4x_4 & \phi_7x_7 \\
                            \phi_2x_2 & \phi_5x_5 & \phi_8x_8 \\
                            \phi_3x_3 & \phi_6x_6 & \phi_9x_9 \\
                          \end{array}
                        \right],
\end{equation}
where $x_i$ are the components of vector $\mathbf{M}_9\mathbf{s}$, $\mathbf{s}$ being the information symbol vector carved from $q$-HEX$^9$ constellation.\smallskip

\subsubsection{Non-vanishing minimum determinant}
By proceeding as previously, we need to determine the coefficients $\phi_l, l=1,\ldots,9$ that guarantee the non-vanishing minimum determinant. In order to get $|\phi_l|=1$ so that a uniform average energy is transmitted per antenna, and to obtain better values of the determinant, we limit the choice of $\phi_l$ to $\phi_l=\pm1, \pm j$.

By developing the code determinant using symbolic computation under Mathematica, we find that it has the same expression as the $3 \times 3$ perfect code determinant by choosing $\phi_l$ as the Fourier matrix coefficients in $\mathbb{Q}(j)$
\begin{equation}
\Phi=\left[
       \begin{array}{ccc}
         1 & 1 & 1 \\
         1 & j & j^2 \\
         1 & j^2 & j \\
       \end{array}
     \right].
\end{equation}
Therefore, the $3\times 3$ infinite code $\bm{\Gamma}(\mathbf{s})$ has non-vanishing minimum determinant equal to
\begin{equation}
\delta_{\min}(\mathcal{C})=\frac{1}{d_{\mathbb{K}_1}}=\frac{1}{49}.
\end{equation}\smallskip

\subsubsection{Delay-tolerance}
On the other hand, to prove the delay-tolerance of this code, we should guarantee that the corresponding shifted codeword matrices are full rank. Therefore, it suffices to verify that for each asynchronous matrix there exists a square $3\times3$ matrix that is full rank \ie its determinant is non-zero. In fact, if we enumerate all the delay profiles, it can be noticed that the problem of guaranteeing full-rank shifted matrices turns to guarantee that
\begin{enumerate}
  \item[-] All component products $\subseteq d_\mathrm{p}$ are non-zero. This condition is always verified since the product distance $d_\mathrm{p}=\prod_{i=1}^{9}|x_i|\neq 0$ over $\mathbb{Z}[j]$ as $d_{\mathrm{p,min}}=\frac{1}{\sqrt{{49}^{3}27^{3}}}$.
  \item[-] All $2\times2$ minors of $\bm{\Gamma(\mathbf{s})}$ are non-zero that is equivalent to verify that the $9$ entries of the cofactor matrix of $\bm{\Gamma}$ are non-zero.
\end{enumerate}

In order to prove the second condition,
we find two unitary matrices $\mathbf{U}$ and $\mathbf{V}$ such that the codeword matrix $\mathbf{\Gamma}$ can be written as $\bm{\Gamma}=\mathbf{U}\mathbf{Z}\mathbf{V}$ for all $\mathbf{s}$, with $\mathbf{Z}$ is the perfect code matrix and $\mathbf{U}$ and $\mathbf{V}$ are defined by
\begin{equation}
\mathbf{U}=
\left[
  \begin{array}{ccc}
    1 & 0 & 0 \\
    0 & j^2\zeta_{9}^2 & 0 \\
    0 & 0 & j^2\zeta_{9} \\
  \end{array}
\right]~,~~
\mathbf{V}=\frac{1}{\sqrt{3}}
\left[
  \begin{array}{ccc}
    1 & 1 & 1 \\
    \zeta_{9} & j\zeta_{9} & j^2\zeta_{9} \\
    \zeta_{9}^{2} & j^2\zeta_{9}^{2} & j\zeta_{9}^{2} \\
  \end{array}
\right].
\end{equation}

Let define the cofactor matrix of the perfect code by $\widetilde{\mathbf{Z}}$. Since $\mathbf{Z}$ is a finite subset of the cyclic division algebra $\mathcal{A}_1$, $\widetilde{\mathbf{Z}}$ is also a subset of $\mathcal{A}_1$ taken from the lattice
$
\Lambda=\mathcal{O}_{\mathbb{K}_1}\oplus e\mathcal{O}_{\mathbb{K}_1}\oplus e^2\mathcal{O}_{\mathbb{K}_1}.
$
with $e^3=j$ and $\mathcal{O}_{\mathbb{K}_1}$ is the ring of integers of $\mathbb{K}_1$. Hence, the cofactor matrix can be represented as a $3\times 3$ codeword matrix. For simplicity, we denote by $\bar{z}=\sigma_1(z)$ and $\bar{\bar{z}}=\sigma_1^2(z)$, the conjugates of an entry of the codeword matrix. The cofactor codeword matrix is then defined by
\begin{equation}
\widetilde{\mathbf{Z}}=
\left[
  \begin{array}{ccc}
    z_{1} & z_{2} & z_{3} \\
    j\bar{z}_{3} & \bar{z}_{1} & \bar{z}_{2} \\
    j\bar{\bar{z}}_{2} & j\bar{\bar{z}}_{3} & \bar{\bar{z}}_{1} \\
  \end{array}
\right],
\end{equation}
where each diagonal $\widetilde{\mathbf{Z}}_i=\mathbf{M}_1[s_i,s_{i+1},s_{i+2}]^T,~ i=1,\ldots,3$.

Since $\bm{\Gamma}=\mathbf{U}\mathbf{Z}\mathbf{V}$, we denote $\widetilde{\bm{\Gamma}}$ its cofactor matrix. It is given by
$\widetilde{\bm{\Gamma}}=\mathbf{V}^{\dag}\widetilde{\mathbf{Z}}\mathbf{U}^{\dag}$ and satisfies
\begin{equation}
\bm{\Gamma}\widetilde{\bm{\Gamma}}=\mathbf{U}\mathbf{Z}\mathbf{V}\mathbf{V}^{\dag}\widetilde{\mathbf{Z}}\mathbf{U}^{\dag}= \det(\mathbf{Z})\mathbf{I},
\end{equation}
with
\begin{equation}
\mathbf{U}^{\dag}=
\left[
  \begin{array}{ccc}
    1 & 0 & 0 \\
    0 & \zeta_{9} & 0 \\
    0 & 0 & \zeta_{9}^2 \\
  \end{array}
\right]~,~~
\mathbf{V}^{\dag}=\frac{1}{\sqrt{3}}
\left[
  \begin{array}{ccc}
    1 & j^2\zeta_{9}^2 & j^2\zeta_{9} \\
    1 & j\zeta_{9}^{2} & \zeta_{9}   \\
    1 & \zeta_{9}^{2} & j\zeta_{9}  \\
  \end{array}
\right].
\end{equation}
Developing the cofactor matrix $\widetilde{\bm{\Gamma}}$, we get
\begin{eqnarray}\label{CofactorMatrix}
\widetilde{\bm{\Gamma}}=\mathbf{V}^{\dag}\widetilde{\mathbf{Z}}\mathbf{U}^{\dag} &=&
   \left[
    \begin{array}{ccc}
      z_{1}+\zeta_{9}^{2}\bar{z}_{3}+\zeta_{9}\bar{\bar{z}}_{2} &
      \zeta_{9}z_{2}+\bar{z}_{1}+\zeta_{9}^{2}\bar{\bar{z}}_{3} &
      \zeta_{9}^{2}z_{3}+\zeta_{9}\bar{z}_{2}+\bar{\bar{z}}_{1} \\

      z_{1}+j^2\zeta_{9}^{2}\bar{z}_{3}+j\zeta_{9}\bar{\bar{z}}_{2} &
      \zeta_{9}z_{2}+j^2\bar{z}_{1}+j\zeta_{9}^{2}\bar{\bar{z}}_{3} & \zeta_{9}^{2}z_{3}+j^2\zeta_{9}\bar{z}_{2}+j\bar{\bar{z}}_{1} \\

      z_{1}+j\zeta_{9}^{2}\bar{z}_{3}+j^2\zeta_{9}\bar{\bar{z}}_{2} &
      \zeta_{9}z_{2}+j\bar{z}_{1}+j^2\zeta_{9}^{2}\bar{\bar{z}}_{3} &
      \zeta_{9}^{2}z_{3}+j\zeta_{9}\bar{z}_{2}+j^2\bar{\bar{z}}_{1} \\
   \end{array}
  \right].
\end{eqnarray}\smallskip
Note that the Galois group $\mathrm{G}=\mathrm{Gal}(\mathbb{K}/\mathbb{F})$ has two generators $\sigma_1$ and $\sigma_2$, it is given by \begin{equation}
\mathrm{G}=\mathrm{Gal}(\mathbb{K}_1/\mathbb{F})\times \mathrm{Gal}(\mathbb{K}_2/\mathbb{F})=\langle\sigma_1,\sigma_2\rangle=\{1, \sigma_1,\sigma_2, \sigma_1^2,\sigma_2^2, \sigma_1\sigma_2,\sigma_1\sigma_2^2,\sigma_1^2\sigma_2,\sigma_1^2\sigma_2^2\}.
\end{equation}
From the expression of $\widetilde{\bm{\Gamma}}$ \eqref{CofactorMatrix}, we define
\begin{eqnarray}
{X}_1\!=\! z_{1}+\zeta_{9}^{2}\bar{z}_{3}+\zeta_{9}\bar{\bar{z}}_{2}, \!&
\sigma_1\sigma_2({X}_1)\! =\! \bar{z}_{1}+j^{2}\zeta_{9}^{2}\bar{\bar{z}}_{3}+j\zeta_{9}z_{2}, \!&
\sigma_1^{2}\sigma_2^2({X}_1)\! =\! \bar{\bar{z}}_{1}+j^{2}\zeta_{9}\bar{z}_{2}+j\zeta_{9}^{2}z_{3}, \nonumber \\
{X}_2 \!=\! \bar{z}_{1}+\zeta_{9}z_{2}+\zeta_{9}^{2}\bar{\bar{z}}_{3}, \!&
\sigma_1\sigma_2({X}_2) \!=\! \bar{\bar{z}}_{1}+j\zeta_{9}\bar{z}_{2}+j^{2}\zeta_{9}^{2}z_{3}, \!&
\sigma_1^{2}\sigma_2^2({X}_{2}) \!=\! z_{1}+j^{2}\zeta_{9}\bar{\bar{z}}_{2}+j\zeta_{9}^{2}\bar{z}_{3}, \nonumber \\
{X}_{3} \!=\! \bar{\bar{z}}_{1}+\zeta_{9}\bar{z}_{2}+\zeta_{9}^{2}z_{3}, \!&
\sigma_1\sigma_2({X}_{3}) \!=\! z_{1}+j\zeta_{9}\bar{\bar{z}}_{2}+j^{2}\zeta_{9}^{2}\bar{z}_{3}, \!&
\sigma_1^{2}\sigma_2^2({X}_{3}) \!=\! \bar{z}_{1}+j^{2}\zeta_{9}z_{2}+j\zeta_{9}^{2}\bar{\bar{z}}_{3}
\end{eqnarray}
the elements ${X}_i$ and their conjugates by the embeddings $(\sigma_1\sigma_2)^k, k=0,\ldots,2$ with $\sigma_1^0\sigma_2^0(X_i)=X_i,\sigma_1^1\sigma_2^1(X_i)=\sigma_1\sigma_2(X_i)$. We also have
${X}_{2} = \sigma_{1}({X}_{1})$ and  ${X}_{3}=\sigma_{1}^{2}({X}_{1})$
the conjugates of ${X}_{1}$ by the embeddings $\sigma_1^k$. Then, the cofactor matrix can be rewritten as
\begin{equation}
    \widetilde{\bm{\Gamma}}=\left[
                                \begin{array}{ccc}
                                  X_1 & X_2 & X_3 \\
                                  \sigma_1\sigma_2(X_3) & j^2\sigma_1\sigma_2(X_1) & j\sigma_1\sigma_2(X_2) \\
                                  \sigma_1^{2}\sigma_2^2(X_2) & j\sigma_1^{2}\sigma_2^2(X_3) & j^2\sigma_1^{2}\sigma_2^2(X_1) \\
                                \end{array}
                              \right]
   = \left[
      \begin{array}{ccc}
        {X}_{1} & \sigma_{1}({X}_{1}) & \sigma_{1}^{2}({X}_{1}) \\
        \sigma_2({X}_{1}) & j^{2}\sigma_1\sigma_2({X}_{1}) & j\sigma_1^2\sigma_2({X}_{1}) \\
        \sigma^{2}_2({X}_{1}) & j\sigma_1\sigma_{2}^{2}({X}_{1}) & j^{2}\sigma_1^2\sigma_2^{2}({X}_{1}) \\
      \end{array}
    \right].
\end{equation}

Finally, computing the product distance of this matrix, we get the product of all the $9$ entries $\tilde{X}_i \in \mathbb{K}$ that is the product of all the $9$ conjugates of $X_1\in \mathbb{K}$, and thus
\begin{equation}
d_{\mathrm{p}}=\prod_{i=1}^9|\tilde{X}_i|=|N_{\mathbb{K}/\mathbb{F}}(X_1)|=\frac{1}{\sqrt{d_{\mathbb{K}}}}=\frac{1}{\sqrt{49^327^3}}.
\end{equation}
As a result, the elements of $\widetilde{\bm{\Gamma}}$ are all non-zero unless $\mathbf{s}=\mathbf{0}$ which concludes our proof on the full-diversity of the $3\times 3$ code $\bm{\Gamma}$, hence its delay-tolerance for any arbitrary delay profile.


\subsection{$4\times4$ Code based on $4\times4$ Perfect Code}

Similarly to the $2\times 2$ case, the $4\times 4$ code is derived over $\mathbb{F}=\mathbb{Q}(i)$ based on the $4\times4$ perfect code algebra. Let $\theta_1=\zeta_{15}+\zeta_{15}^{-1}=2\cos(\frac{2\pi}{15})$, the relative extension is $\mathbb{K}_1=\mathbb{Q}(i,\theta_1)$ of degree $[\mathbb{K}_1:\mathbb{F}]=4$ and its relative discriminant is $d_{\mathbb{K}_1}=1125$. The cyclic Galois group $\mathbb{K}_1/\mathbb{F}$ is generated by $\sigma_1:\zeta_{15}+\zeta_{15}^{-1}\mapsto \zeta_{15}^2+\zeta_{15}^{-2}$. The integral basis is
$
\mathrm{B}_1=\{v_k\}_{k=1}^4 = \{(1-3i)+i\theta_1^2,(1-3i)\theta_1+i\theta_1^3,
-i+(-3+4i)\theta_1+(1-i)\theta_1^3,(-1+i)-3\theta_1+\theta^2_1+\theta_1^3\}
$
and the complex rotated lattice of $\mathbb{Z}[i]^4$ is generated by the unitary matrix
\begin{equation}
\mathbf{M_1}=\frac{1}{\sqrt{15}}\left[
               \begin{array}{cccc}
                 v_1 & v_2 & v_3 & v_4 \\
                 \sigma_1(v_1) & \sigma_1(v_2) & \sigma_1(v_3) & \sigma_1(v_4) \\
                 \sigma_1^2(v_1) & \sigma_1^2(v_2) & \sigma_1^2(v_3) & \sigma_1^2(v_4) \\
                 \sigma_1^3(v_1) & \sigma_1^3(v_2) & \sigma_1^3(v_3) & \sigma_1^3(v_4) \\
               \end{array}
             \right].
\end{equation}

The second 
relative extension $\mathbb{K}_2$ is chosen such that its degree is $4$ over $\mathbb{F}$ and has coprime discriminant with $\mathbb{K}_1$. Let $\mathbb{K}_2=\mathbb{Q}(\zeta_{16})$ this cyclotomic extension with $d_{\mathbb{K}_2}=256$ and $\zeta_{16}=e^{i\pi/8}$ the primitive $16^{th}$ root of unity. The cyclic Galois group is generated by $\sigma_2:\zeta_{16}\mapsto i\zeta_{16}$. The integral basis of $\mathbb{K}_2$ is $\mathrm{B}_2=(1,\zeta_{16},\zeta_{16}^2,\zeta_{16}^3)$ and the lattice generator matrix in $\mathbb{Z}[i]^4$ is given by
\begin{equation}
\mathbf{M_2}=\frac{1}{2}\left[
               \begin{array}{cccc}
                 1 & \zeta_{16} & \zeta_{16}^2 & \zeta_{16}^3 \\
                 1 & i\zeta_{16} & -\zeta_{16}^2 & -i\zeta_{16}^3 \\
                 1 & -\zeta_{16} & \zeta_{16}^2 & -\zeta_{16}^3 \\
                 1 & -i\zeta_{16} & -\zeta_{16}^2 & i\zeta_{16}^3 \\
               \end{array}
             \right].
\end{equation}

Then, the tensor product of both cyclic extensions defines the compositum field  $\mathbb{K}=\mathbb{K}_1\mathbb{K}_2=\mathbb{Q}(i,2\cos(\frac{2\pi}{15}),\zeta_{16})$ of order $16$ over $\mathbb{Q}(i)$. Accordingly, the $16$-dimensional complex lattice is generated by the $16 \times 16$ unitary matrix $\mathbf{M}_{16}=\mathbf{M}_{2}\otimes\mathbf{M}_{1}$.
The $16$ codeword elements are derived from the linear combination $\mathbf{M}_{16}\mathbf{s}$ of $q$-QAM information symbols. They are then distributed in the $4\times 4$ codeword matrix and assigned the coefficients $\phi_l,l=1,\ldots,16$ as
\begin{equation}
\bm{\Gamma}(\mathbf{s})=\left[
                          \begin{array}{cccc}
                            \phi_1x_1 & \phi_5x_5 & \phi_9x_9 & \phi_{13}x_{13}\\
                            \phi_2x_2 & \phi_6x_6 & \phi_{10}x_{10} & \phi_{14}x_{14} \\
                            \phi_3x_3 & \phi_7x_7 & \phi_{11}x_{11} & \phi_{15}x_{15} \\
                            \phi_4x_4 & \phi_8x_8 & \phi_{12}x_{12} & \phi_{16}x_{16} \\
                          \end{array}
                        \right].
\end{equation}\smallskip

\subsubsection{Non-vanishing minimum determinant}

The coefficients $\phi_l$ are restricted to $|\phi_l|=1$ for uniform energy transmission and should satisfy the NVD criterion. Therefore, as in previous dimensions, computing the code determinant using symbolic computation under Mathematica, we find that such coefficients corresponding to the Fourier matrix coefficients in $\mathbb{Q}(i)$ allow to get a $4\times 4$ space-time code with the same determinant as the perfect code. We have
\begin{equation}
\Phi=\left[
       \begin{array}{cccc}
         1 & 1 & 1 & 1 \\
         1 & i & -1 & -i \\
         1 & -1 & 1 & -1 \\
         1 & -i & -1 & i \\
       \end{array}
     \right].
\end{equation}
Therefore, the $4\times 4$ infinite code $\bm{\Gamma}(\mathbf{s})$ has non-vanishing minimum determinant
\begin{equation}
\delta_{\min}(\mathcal{C})=\frac{1}{d_{\mathbb{K}_1}}=\frac{1}{1125},
\end{equation}
and the $4\times4$ codeword matrix is defined for $x_1=X$ by
\begin{equation}\label{4x4CodewordMatrix}
\mathbf{\Gamma(s)}=\left[
                     \begin{array}{cccc}
                       x_1 & x_5 & x_9 & x_{13} \\
                       x_2 & ix_6 & -x_{10} & -ix_{14} \\
                       x_3 & -x_7 & x_{11} & -x_{15} \\
                       x_4 & -ix_8 & -x_{12} & ix_{16} \\
                     \end{array}
                   \right]
=\left[
   \begin{array}{cccc}
     X & \sigma_2(X) & \sigma_2^{2}(X) & \sigma_2^{3}(X) \\
     \sigma_1(X) & i\sigma_2\sigma_1(X) & -\sigma_2^{2}\sigma_1(X) & -i\sigma_2^{3}\sigma_1(X) \\
     \sigma_1^{2}(X) & -\sigma_2\sigma_1^{2}(X) & \sigma_2^{2}\sigma_1^{2}(X) & -\sigma_2^{3}\sigma_1^{2}(X) \\
     \sigma_1^{3}(X) & -i\sigma_2\sigma_1^{3}(X) & -\sigma_2^{2}\sigma_1^{3}(X) & i\sigma_2^{3}\sigma_1^{3}(X) \\
   \end{array}
 \right].
\end{equation}\smallskip

\subsubsection{Delay-tolerance}

Now, let us examine the delay-tolerance aspect of this code. For this task, we start by enumerating all the types of delay profiles. Consider the integer numbers $a,b,c,d$ with $a\neq b \neq c \neq d$ and $ 0\leq a,b,c,d \leq 3$, we can define four types of profiles as:
\begin{enumerate}
    \item[-] Type $1$ of form $\bm{\mathfrak{d}}=(a,b,c,d)$  
    \item[-] Type $2$ of form $\bm{\mathfrak{d}}=(a,a,b,c)$ 
    \item[-] Type $3$ of form $\bm{\mathfrak{d}}=(a,a,b,b)$ 
    \item[-] Type $4$ of form $\bm{\mathfrak{d}}=(a,a,a,b)$ 
\end{enumerate}


Each of the asynchronous shifted codeword matrices corresponding to these profiles is full rank if and only if it includes a square $4\times4$ matrix that is full rank \ie a $4\times4$ minor that is non-zero. This will be proved in the sequel for the different delay profile types.\bigskip

\textbf{Types $1$ and $4$}\smallskip

If we consider the delay profiles of types $1$ and $4$, for instance $\bm{\mathfrak{d}}_1=(0,1,2,3)$, $\bm{\mathfrak{d}}_2=(0,0,0,1)$ and $\bm{\mathfrak{d}}_3=(3,0,0,0)$, the $4\times4$ minors $\mathcal{M}_{4\times4}$ relative to the $4\times4$ shifted matrices can have one of these expressions
\begin{enumerate}
\item[-] The product of some components of the codeword matrix $\mathbf{\Gamma}$:  $\prod\limits_{1\leq l\leq 16}|\phi_lx_l|$
\item[-] The product of one component and a $3\times 3$ minor $\mathcal{M}_{3\times3}$: $\phi_lx_l\mathcal{M}_{3\times3}$
\end{enumerate}
%


\underline{\emph{Proof 1}}

In the first case, we have by construction that all component products $\subseteq d_\mathrm{p}=\prod\limits_{k=1}^{16}|x_k|$ are non-zero since $d_{\mathrm{p,min}}=\frac{1}{\sqrt{d_{\mathbb{K}}}}=\frac{1}{\sqrt{{1125}^{4}256^{4}}}$.

In the second case, following the same analysis of the $3\times3$ space-time code, we find the unitary matrices $\mathbf{U}$ and $\mathbf{V}$
\begin{equation}
\mathbf{U}=\left[
            \begin{array}{cccc}
             1 & 0 & 0 & 0 \\
             0 & -i\zeta_{16}^{3} & 0 & 0 \\
             0 & 0 & -i\zeta_{16}^2 & 0 \\
             0 & 0 & 0 & -i\zeta_{16}
           \end{array}
        \right]~,~~
\mathbf{V}=\frac{1}{2}
\left[\begin{array}{cccc}
        1 & 1 & 1 & 1 \\
        \zeta_{16} & i\zeta_{16} & -\zeta_{16} & -i\zeta_{16} \\
        \zeta_{16}^2 & -\zeta_{16}^2 & \zeta_{16}^2 & -\zeta_{16}^2 \\
        \zeta_{16}^{3} & -i\zeta_{16}^{3} & -\zeta_{16}^{3} & i\zeta_{16}^{3}
      \end{array}
\right]~,
\end{equation}
such that the new $4\times4$ code $\bm{\Gamma}$ can be written as $\bm{\Gamma}=\mathbf{U}\mathbf{Z}\mathbf{V}$, $\mathbf{Z}$ being the $4\times4$ perfect code. Then, we derive the cofactor matrix $\widetilde{\bm{\Gamma}}$ and prove that it has non-zero entries as its product distance is non-zero. Thus, the $3\times3$ minors are full-rank yielding full-rank shifted matrices.\bigskip

\textbf{Type $2$}\smallskip

For delay profiles of type $2$, for instance $\bm{\mathfrak{d}}_1=(0,0,1,3)$, $\bm{\mathfrak{d}}_2=(2,2,1,0)$ and $\bm{\mathfrak{d}}_3=(3,3,0,2)$, we can find $4\times 4$ minors $\mathcal{M}_{4\times4}$ in the relative shifted codeword matrices that are equal to
\begin{equation}
\mathcal{M}_{4\times4}=(\phi_jx_j)(\phi_kx_k)\mathcal{M}_{2\times2},
\end{equation}
where $\mathcal{M}_{2\times2}$ has its components $\phi_lx_l$ such that only one $\phi_l=\pm i$. So, the $4\times4$ minors are non-zero if these $2\times2$ minors are non-zero for any $\mathbf{s}\neq\mathbf{0}$.\smallskip

\underline{\emph{Proof 2}}

Let
\begin{equation}
\mathcal{M}_1=
\left|
\begin{array}{cc}
  x_1 & x_5 \\
  x_2 & ix_6 \\
\end{array}
\right|
\end{equation}
be such $2\times2$ minor and consider any $3\times3$ minor $\mathcal{M}_{3\times3}$ that includes $\mathcal{M}_1$, for example
\[
\mathcal{M}_{3\times3}=
\left|
  \begin{array}{ccc}
    x_1 & x_5 & x_9 \\
    x_2 & ix_6 & -x_{10} \\
    x_3 & -x_7 & x_{11} \\
  \end{array}
\right|.
\]
It can be expanded into
\begin{eqnarray}
\mathcal{M}_{3\times3} &=& x_{11}\left|
                                \begin{array}{cc}
                                  x_1 & x_5 \\
                                  x_2 & ix_6 \\
                                \end{array}
                              \right|+
                               x_{7}\left|
                                \begin{array}{cc}
                                  x_1 & x_9 \\
                                  x_2 & -x_{10} \\
                                \end{array}
                              \right|+
                               x_{3}\left|
                                \begin{array}{cc}
                                  x_5 & x_9 \\
                                  ix_6 & -x_{10} \\
                                \end{array}
                              \right|\nonumber \\
                    &=& x_{11}\mathcal{M}_1+x_7\mathcal{M}_2+x_3\mathcal{M}_3
\end{eqnarray}
By developing the $2\times2$ minors, we have according to the $4\times4$ codeword matrix \eqref{4x4CodewordMatrix}
\begin{eqnarray}
\mathcal{M}_1 &=& iX\sigma_2\sigma_1(X)-\sigma_1(X)\sigma_2(X) \nonumber\\
\mathcal{M}_2 &=& -X\sigma^2_2\sigma_1(X)-\sigma_1(X)\sigma_2^2(X)= -Y-\sigma^2_2(Y) \nonumber\\
\mathcal{M}_3 &=& -\sigma_2(X)\sigma_2^2\sigma_1(X)-i\sigma_2\sigma_1(X)\sigma_2^2(X)=i\sigma_2(\mathcal{M}_1)
\end{eqnarray}
then
\begin{equation}
\mathcal{M}_{3\times3}=\sigma_2^2\sigma_1^2(X)\mathcal{M}_1+\sigma_2\sigma^2_1(X)\mathcal{M}_2+i\sigma_1^2(X)\sigma_2(\mathcal{M}_1).
\end{equation}
If $\mathcal{M}_1= 0$,  $\mathcal{M}_{3\times3}=\sigma_2\sigma^2_1(X)\mathcal{M}_2$ can be zero since $\mathcal{M}_2$ is a trace and can be zero (if $Y=\gamma\zeta_{16}, \gamma\in \mathbb{F}_1, Y \in \mathbb{K}$). However, we have from \emph{Proof 1} that any $3\times3$ minor is non-zero over $\mathbb{Z}[i]$. Thus, $\mathcal{M}_1$ cannot be zero over $\mathbb{Z}[i]$ unless $\mathbf{s}=\mathbf{0}$. By a similar analysis, we can prove that any $2\times2$ minor of the same form of $\mathcal{M}_1$ is non-zero for $\mathbf{s}\neq\mathbf{0}$. \bigskip

\textbf{Type $3$}\smallskip

For this type, we distinguish two cases of profiles:
\begin{enumerate}
\item[3I-] $a=2,3, b=0$ such as $\bm{\mathfrak{d}}_1=(2,2,0,0), \bm{\mathfrak{d}}_2=(3,3,0,0)$
\item[3II-] $a=1, b=0$ such as $\bm{\mathfrak{d}}=(1,1,0,0)$
\end{enumerate}

In the first case, there exist $4\times4$ minors that are equal to the product of two $2\times2$ minors such that these $\mathcal{M}_{2\times2}$ have there components $\phi_lx_l$ with only one $\phi_l=\pm i$, hence are non-zero according to \emph{Proof 2}.

In the second case, the $4\times4$ minors are functions of $2\times2$ minors $\mathcal{M}_{2\times2}$ as following
\begin{equation}
\mathcal{M}_{4\times4}=\sum\limits_{k=0}^2\prod\limits_{l=0}^1\mathcal{M}_{2\times2, k+l+1},
\end{equation}
and thus we have to prove that this sum is non-zero over $\mathbb{Z}[i]$. For this task and without loss of generality, we consider the delay profile $\bm{\mathfrak{d}}=(1,1,0,0)$.

\underline{\emph{Proof 3}}

Let the $4\times4$ minor relative to this delay profile be
\begin{eqnarray}
\mathcal{M}_{4\times4}&=&\!\!\left|
                         \begin{array}{cccc}
                           0 & x_1 & x_5 & x_9 \\
                           0 & x_2 & ix_6 & -x_{10} \\
                           x_3 & -x_7 & x_{11} & -x_{15} \\
                           x_4 & -ix_8 & -x_{12} & ix_{16} \\
                         \end{array}
                       \right| \nonumber\\
&=&\!\!\left|
     \begin{array}{cc}
       x_1 & x_5 \\
       x_2 & ix_6 \\
     \end{array}
   \right|\left|
            \begin{array}{cc}
              x_3 & -x_{15} \\
              x_4 & ix_{16} \\
            \end{array}
          \right|-
\left|
     \begin{array}{cc}
       x_1 & x_9 \\
       x_2 & -x_{10} \\
     \end{array}
   \right|\left|
            \begin{array}{cc}
              x_3 & x_{11} \\
              x_4 & -x_{12} \\
            \end{array}
          \right|+
\left|
     \begin{array}{cc}
       x_5 & x_9 \\
       ix_6 & -x_{10} \\
     \end{array}
   \right|\left|
            \begin{array}{cc}
              x_3 & -x_7 \\
              x_4 & -ix_8 \\
            \end{array}
          \right| \nonumber \\
&=&\!\!\mathcal{M}_1\mathcal{M}_2-\mathcal{M}_3\mathcal{M}_4+\mathcal{M}_5\mathcal{M}_6,
\end{eqnarray}
with according to the codeword matrix in Equation \eqref{4x4CodewordMatrix}
\begin{eqnarray}
&\mathcal{M}_1=iX\sigma_2\sigma_1(X)-\sigma_1(X)\sigma_2(X)~~~~~~~~
&\mathcal{M}_2=-i\sigma_2^3\sigma_1^2(\mathcal{M}_1) \nonumber\\
&\mathcal{M}_3=-X\sigma_2^2\sigma_1(X)-\sigma_1(X)\sigma_2^2(X)~~~~~~~~
&\mathcal{M}_4=\sigma_1^2(\mathcal{M}_3) \nonumber\\
&\mathcal{M}_5=i\sigma_2(\mathcal{M}_1)~~~~~~~~~~~~~~~~~~~~~~~~~~~~~~~
&\mathcal{M}_6=-\sigma_1^2(\mathcal{M}_1)
\end{eqnarray}
then
\begin{equation}
\mathcal{M}_{4\times4}=-i\mathcal{M}_1\sigma_2^3\sigma_1^2(\mathcal{M}_1)-\mathcal{M}_3\sigma_1^2(\mathcal{M}_3)-i\sigma_2(\mathcal{M}_1)\sigma_1^2(\mathcal{M}_1).
\end{equation}
By denoting the first term in this expression $\mathcal{P}_1$ and the second term $\mathcal{P}_2$, then
\begin{equation}
\mathcal{M}_{4\times4}=\mathcal{P}_1+\mathcal{P}_2+\sigma_2(\mathcal{P}_1).
\end{equation}
Recalling $\mathcal{M}_3$, we can notice that it can be written as
\[
\mathcal{M}_3=-Y-\sigma_2^2(Y)=-\mathrm{Tr}_{\mathbb{K}/\mathbb{F}_1}(Y)~~~\mathrm{with}~~~Y=X\sigma_2^2\sigma_1(X)~\in~\mathbb{K}.
\]
Let $Y \in \mathbb{K}$ be $Y = A_1+\zeta_{16}B_1 +\zeta_{16}^2C_1+\zeta_{16}^3D_1$ with $A_1,B_1,C_1,D_1 \in \mathbb{K}_1$. Then,
\begin{equation}
\mathcal{M}_3=-2A_1-2\zeta_{16}^2C_1=a_1+\zeta_8b_1~~~\mathrm{with}~~~a_1,b_1\in \mathbb{K}_1.
\end{equation}
For simplicity, we denote the conjugate $\sigma_1(x)=\bar{x}$, so we have
\begin{equation}
\mathcal{P}_2=-\mathcal{M}_3\sigma_1^2(\mathcal{M}_3)=-(a_1+\zeta_8b_1)(\bar{\bar{a}}_1+\zeta_8\bar{\bar{b}}_1)=-a_1\bar{\bar{a}}_1-\zeta_8^2b_1\bar{\bar{b}}_1-\zeta_8(a_1\bar{\bar{b}}_1+\bar{\bar{a}}_1b_1).
\end{equation}

\begin{figure} [h]%
\centering %
\includegraphics[width = 0.8\textwidth]{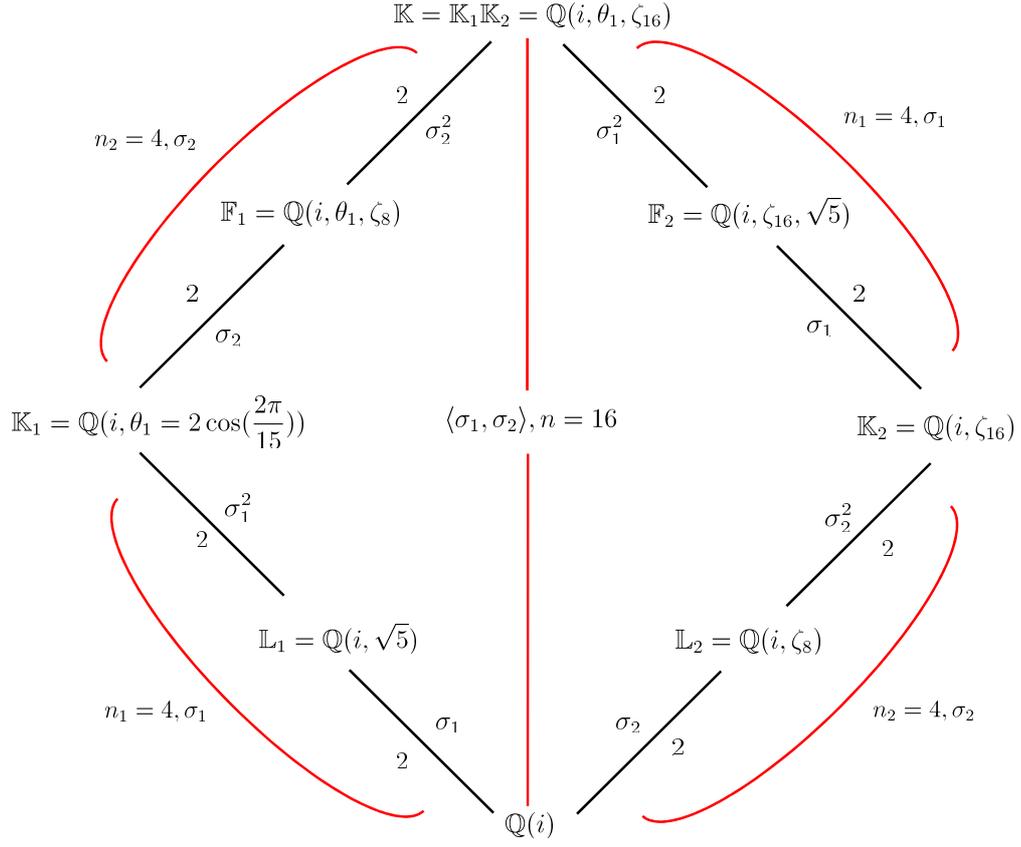} \vspace{-8pt}%
\caption{Nested Sequences of fields} %
\label{NestedSeq}%
\end{figure}

Let us now examine the nested sequences of fields included in the compositum field $\mathbb{K}$ in Figure \ref{NestedSeq}. We have
\begin{eqnarray}
& \mathbb{F}=\mathbb{Q}(i) \subset \mathbb{L}_1=\mathbb{Q}(i,\sqrt{5}) \subset \mathbb{K}_1=\mathbb{Q}(i,\theta_1) \subset \mathbb{F}_1=\mathbb{Q}(i,\theta_1,\zeta_8) \subset \mathbb{K}=\mathbb{Q}(i,\theta_1,\zeta_{16}) & \\
& \mathbb{F}=\mathbb{Q}(i) \subset \mathbb{L}_2=\mathbb{Q}(i,\zeta_8) \subset \mathbb{K}_2=\mathbb{Q}(i,\zeta_{16}) \subset \mathbb{F}_2=\mathbb{Q}(i,\zeta_{16},\sqrt{5}) \subset \mathbb{K}=\mathbb{Q}(i,\theta_1,\zeta_{16}) &
\end{eqnarray}
with the perfect algebra $\mathcal{PA}=(\mathbb{K}_1/\mathbb{L}_1, \sigma_1^2, \gamma_1)=\mathbb{K}_1\oplus u_1\mathbb{K}_1$, where $u_1^2=\gamma_1=i$ and $\sigma_1: \theta_1= \zeta_{15}+\zeta_{15}^{-1} \mapsto \theta_1^2-2 =\zeta_{15}^2+\zeta_{15}^{-2}, \sqrt{5} \mapsto -\sqrt{5}$. As we have $\sigma_1(\sqrt{5})=-\sqrt{5}$, $\mathbb{L}_1$ is the subfield fixed by $\langle\sigma_1^2\rangle$ the subgroup of order $[\mathbb{L}_1 : \mathbb{F}]=2$ of the Galois group Gal$(\mathbb{K}_1/\mathbb{F})=\langle\sigma_1\rangle$ \cite{Hollanti:2008-2}.

On the other hand, we have the cyclotomic algebra $\mathcal{CA}=(\mathbb{K}_2/\mathbb{L}_2, \sigma_2^2: \zeta_{16}\mapsto -\zeta_{16}, \gamma_2=1+\zeta_8)=\mathbb{K}_2\oplus u_2\mathbb{K}_2$, and $\sigma_2: \zeta_{16} \mapsto i\zeta_{16}, \zeta_8 \mapsto -\zeta_8$. As we have $\sigma_2(\zeta_8)=-\zeta_8$, $\mathbb{L}_2$ is the subfield fixed by $\langle\sigma_2^2\rangle$ the subgroup of order $[\mathbb{L}_2 : \mathbb{F}]=2$ of the Galois group Gal$(\mathbb{K}_2/\mathbb{F})=\langle\sigma_2\rangle$ \cite{Hollanti:2008-2}.

From the nested sequence of fields $(71)$, we can deduce that
\begin{equation}\label{P2}
\mathcal{P}_2=\underbrace{\underbrace{-N_{\mathbb{K}_1/\mathbb{L}_1}(a_1)}\limits_{\in~\mathbb{L}_1}-i\underbrace{N_{\mathbb{K}_1/\mathbb{L}_1}(b_1)}\limits_{\in~\mathbb{L}_1}}\limits_{\in~\mathbb{L}_1~\subset~\mathbb{K}_1}
-\underbrace{\underbrace{\zeta_8}\limits_{\in~\mathbb{F}_1}\underbrace{\mathrm{Tr}_{\mathbb{K}_1/\mathbb{L}_1}(a_1\bar{\bar{b}}_1)}\limits_{\in~\mathbb{L}_1}}\limits_{\in~\mathbb{F}_1~\subset~\mathbb{K}}.
\end{equation}

On the other hand, we have $\mathcal{P}_1 \in \mathbb{K}$, we can define it as $\mathcal{P}_1 = a_2+\zeta_{16}b_2 +\zeta_{16}^2c_2+\zeta_{16}^3d_2$ with $a_2,b_2,c_2,d_2 \in \mathbb{K}_1$, then
\begin{equation}\label{P1}
\mathcal{P}_1+\sigma_2(\mathcal{P}_1)=\underbrace{2a_2}\limits_{\in~\mathbb{K}_1~\subset~\mathbb{K}}+\underbrace{\zeta_{16}(1+i)b_2}\limits_{\in~\mathbb{K}}+\underbrace{\zeta_{16}^3(1-i)d_2}\limits_{\in~\mathbb{K}}.
\end{equation}

Therefore, we can define $\mathcal{M}_{4\times4}$ as
\begin{eqnarray}
\mathcal{M}_{4\times4} &= &\!\! \mathcal{P}_1+\sigma_2(\mathcal{P}_1)+\mathcal{P}_2 \nonumber\\
&=&\!\! (-N_{\mathbb{K}_1/\mathbb{L}_1}(a_1)-iN_{\mathbb{K}_1/\mathbb{L}_1}(b_1)+2a_2)+ \zeta_{16}(1+i)b_2 -\zeta_{16}^2 \mathrm{Tr}_{\mathbb{K}_1/\mathbb{L}_1}(a_1\bar{\bar{b}}_1) + \zeta_{16}^3(1-i)d_2 \nonumber\\
&=&\!\! A+ \zeta_{16}B+\zeta_{16}^2C+\zeta_{16}^3D
\end{eqnarray}
with
\begin{eqnarray}
A &=& -N_{\mathbb{K}_1/\mathbb{L}_1}(a_1)-iN_{\mathbb{K}_1/\mathbb{L}_1}(b_1)+2a_2 ~\in ~ \mathbb{K}_1 \nonumber\\
B &=& (1+i)b_2 ~\in~\mathbb{K}_1 \nonumber\\
C &=& \mathrm{Tr}_{\mathbb{K}_1/\mathbb{L}_1}(a_1\bar{\bar{b}}_1) ~\in~\mathbb{L}_1~\subset~\mathbb{K}_1 \nonumber\\
D &=& (1-i)d_2 ~\in~\mathbb{K}_1
\end{eqnarray}

It can be seen as a vector space of $\mathbb{K}_1$ with basis $(1, \zeta_{16},\zeta_{16}^2,\zeta_{16}^3)$, and thus $\mathcal{M}_{4\times4}=0$ if and only if $A=B=C=D=0$. This condition reduces to
\begin{equation}
\left\{
  \begin{array}{ll}
    A=0 ~ \Rightarrow & \hbox{$N_{\mathbb{K}_1/\mathbb{L}_1}(a_1)+iN_{\mathbb{K}_1/\mathbb{L}_1}(b_1)=2a_2$} ~~~~(i)\\
    B=0 ~ \Rightarrow & \hbox{$b_2=0$} ~~~~(ii) \\
    C=0 ~ \Rightarrow & \hbox{$\mathrm{Tr}_{\mathbb{K}_1/\mathbb{L}_1}(a_1\bar{\bar{b}}_1)=0$} ~~~~(iii)\\
    D=0 ~ \Rightarrow & \hbox{$d_2=0$}~~~~(iv)
  \end{array}
\right.
\end{equation}

So in order to prove that the $4\times4$ minor is non-zero, we have to prove that the latter condition cannot be verified. We proceed by contradiction.

For this task, we show that by assuming that $(ii), (iii), (iv)$ are verified, we cannot have $(i)$. In fact, if $A=0$, one particular case would be when $a_1=b_1=0$, so that $a_2=0$.

\noindent However, if $a_1=b_1=0$ and according to Equations \eqref{P2} and $(iii)$, we have $\mathcal{P}_2=\mathrm{Tr}_{\mathbb{K}_1/\mathbb{F}_1}(Y)=0$. Consider the general case where $X \in \mathbb{K}$, we can define it by
\begin{equation}
X=\alpha+\zeta_{16}\beta ~~~\mathrm{with}~~~\alpha,\beta \in \mathbb{F}_1,
\end{equation}
and its conjugate by $\sigma_1(X)=\bar{\alpha}+\zeta_{16}\bar{\beta}$. We have then
\begin{equation}
Y=X\sigma_2^2\sigma_1(X)=(\alpha+\zeta_{16}\beta)(\bar{\alpha}-\zeta_{16}\bar{\beta})=(\alpha\bar{\alpha}-\zeta_{16}^2\beta\bar{\beta})+\zeta_{16}(\bar{\alpha}\beta-\zeta_{16}^2\beta\bar{\beta}).
\end{equation}
Since $\mathrm{Tr}_{\mathbb{K}/\mathbb{F}_1}(Y)=0$, thus $Y \in \mathbb{K}$ is of the form $Y=\gamma\zeta_{16}$, with $\gamma \in \mathbb{F}_1$. Therefore, we have
\begin{equation}
\alpha\bar{\alpha}-\zeta_{16}^2\beta\bar{\beta}=0 ~\Rightarrow~\alpha\bar{\alpha}=\zeta_{8}\beta\bar{\beta}.
\end{equation}

Let us now compute $\mathcal{P}_1+\sigma_2(\mathcal{P}_1)$ given this condition according to $\mathcal{P}_2=0$. Recall that $\mathcal{P}_1=-i\mathcal{M}_1\sigma_2^3\sigma_1^2(\mathcal{M}_1)$ with $\mathcal{M}_1=iX\sigma_2\sigma_1(X)-\sigma_1(X)\sigma_2(X)$, then $\mathcal{M}_1$ and $\mathcal{P}_1$ can be reduced to
\begin{eqnarray}
\mathcal{M}_1 & = & -2\alpha\bar{\alpha}-2\zeta_{16}\alpha\bar{\beta}=-2\alpha(\bar{\alpha}+\zeta_{16}\bar{\beta})\\
\mathcal{P}_1 & = & -4\alpha\bar{\bar{\alpha}}(\bar{\alpha}+\zeta_{16}\bar{\beta})(\bar{\bar{\bar{\alpha}}}-i\zeta_{16}\bar{\bar{\bar{\beta}}})
\end{eqnarray}
and
\begin{equation}
\mathcal{P}_1+\sigma_2(\mathcal{P}_1)=-4i\alpha\bar{\bar{\alpha}}\Big(2\bar{\alpha}\bar{\bar{\bar{\alpha}}}+\zeta_{16}(1-i)\bar{\alpha}\bar{\bar{\bar{\beta}}}+ \zeta_{16}(1+i)\bar{\bar{\bar{\alpha}}}\bar{\beta} \Big).
\end{equation}

On the other hand, we have according to $(ii), (iv) (b_2=d_2=0)$ and Equation \eqref{P1} that $\mathcal{P}_1+\sigma_2(\mathcal{P}_1)=2a_2$. So, it can be simplified to
\begin{equation}
\mathcal{P}_1+\sigma_2(\mathcal{P}_1)=-8i\alpha\bar{\alpha}\bar{\bar{\alpha}}\bar{\bar{\bar{\alpha}}}.
\end{equation}
Therefore,
\begin{equation}
\mathcal{P}_1+\sigma_2(\mathcal{P}_1)=0 ~\Rightarrow~ \alpha=0.
\end{equation}
However, $\alpha=0$ means that $\mathcal{M}_1=0$ as well. But, we have already proved in \textit{Proof 2} that $\mathcal{M}_1\neq0$ over $\mathbb{Z}[i]$ unless $\mathbf{s}=0$. Consequently, $\mathcal{P}_1+\sigma_2(\mathcal{P}_1)\neq 0$, then $a_2\neq 0$. So Given $a_1=b_1=0$, we prove here that $A\neq0$ and thus, $\mathcal{M}_{4\times4}$ cannot be zero over $\mathbb{Z}[i]$ for $\mathbf{s}\neq 0$.\\

This last proof concludes the analysis on the full-rank asynchronous codeword matrices for the different types of delay profiles \ie the full-diversity of the $4\times 4$ code $\bm{\Gamma}$, hence its delay-tolerance for arbitrary delay profiles.


\section{New Delay-Tolerant Codes from other Perfect Codes}
We derive now delay-tolerant codes from the perfect codes presented in \cite{Elia:2007}. These latter codes differ from the previous ones by the construction of their generator matrices and their non-norm element $\gamma_1$. Whereas this element was chosen as a root of unity in $2,3,4$-dimensional perfect codes $(\gamma_1=i,j)$, for the current codes it is of the form
$$
\gamma_1=\frac{\pi}{\pi^{*}}
$$
where $\pi$ is an element of $\mathbb{K}_1$ and $\pi^{*}$ its complex conjugate. $\pi$ is chosen as a suitable prime in $\mathbb{Z}[i]$ or $\mathbb{Z}[j]$ so that the element $\gamma_1$ is of unit norm and it is non-norm for the extension $\mathbb{K}_1/\mathbb{F}$.

Based on the same approach in Section (\ref{sec732}), the delay-tolerant code is constructed using the tensor product of two number fields with the same degree and coprime discriminants. In previous dimensions, the second field corresponds to the cyclotomic extension $\mathbb{K}_2=\mathbb{Q}(\zeta^M)$  where $\zeta^M$ is the $M^{th}$ root of unity since the non-norm element of the perfect code is itself a root of unity. Consequently, the relative extension will be here $\mathbb{K}_2=\mathbb{Q}(\theta_2)$ with $\theta_2=\sqrt[M]{\gamma_1}$ is the $M^{th}$ root of the non-norm element $\gamma_1$.

\subsection{$2\times2$ Code}
We consider the case of $2$ antennas. The corresponding $2\times2$ perfect code was constructed in \cite{Elia:2007} on the field $\mathbb{F}=\mathbb{Q}(i)$, and thus transmits $q$-QAM symbols. Let $\theta_1=2\cos(\frac{2\pi}{5})$ and $\mathbb{K}_1=\mathbb{Q}(i,\theta_1)$ the relative extension of $\mathbb{F}$ of degree $[\mathbb{Q}(i,\theta_1):\mathbb{Q}(i)]=2$. The cyclic group $\mathbb{K}_1/\mathbb{F}$ is generated by $\sigma_1$ and the cyclic algebra is then $\mathcal{A}_1(\mathbb{K}_1/\mathbb{F},\sigma_1,\gamma_1)$ with the non-norm element $$\gamma_1=\frac{3+2i}{2+3i}$$
The rotated lattice $\mathbb{Z}[i]^2$ is obtained by a technique presented in \cite{Elia:2007, Bayer:2004} different from the one used for previous perfect codes. The generator matrix is numerically given by \cite{Rotations}
\begin{equation}
\mathbf{M}_1=\left[
               \begin{array}{cc}
                 -0.52573 & -0.85065 \\
                 -0.85065 & 0.52573 \\
               \end{array}
             \right].
\end{equation}

Now, let $\mathbb{K}_2=\mathbb{Q}(\theta_2)$ the cyclotomic extension of degree $2$ of $\mathbb{F}$ with $\theta_2=(\gamma_1)^{1/2}$. Its relative discriminant is $d_{\mathbb{K}_2}=52$ and is coprime to $d_{\mathbb{K}_1}=5$. The cyclic Galois group generator is $\sigma_2: \theta_2\mapsto -\theta_2$ and the integral basis is $\mathrm{B}_2=(1,\theta_2)$. The rotated $\mathbb{Z}[i]^2$ lattice is generated by the unitary matrix
\begin{equation}
\mathbf{M}_2=\frac{1}{\sqrt{2}}\left[
               \begin{array}{cc}
                 1 & \theta_2 \\
                 1 & -\theta_2\\
               \end{array}
             \right].
\end{equation}

Then, the compositum of both cyclic extensions is defined by $\mathbb{K}=\mathbb{K}_1\mathbb{K}_2=\mathbb{Q}(i,\theta_1,(\gamma_1)^{1/5})$ of order $4$ over $\mathbb{Q}(i)$ and accordingly the $4$-dimensional complex lattice is generated by the $4 \times 4$ unitary matrix
$
\mathbf{M}_{4}=\mathbf{M}_2\otimes \mathbf{M}_1.
$
The codeword components are derived from the linear combination $\mathbf{x}=\mathbf{M}_{4}\mathbf{s}$ of $q$-QAM information symbols. They are then distributed in the $2\times 2$ codeword matrix assigned by the Fourier matrix coefficients $\phi_l, l=1,\ldots,4$ in dimension $2$ in order to guarantee the same NVD as the corresponding $2\times 2$ perfect code.
%
Both matrices $\mathbf{U}$ and $\mathbf{V}$ can also be derived by replacing $\zeta_8$ by $\theta_2$. Moreover, the code construction allows to have a non-zero product distance, yielding a delay-tolerant code that maintains its full diversity regardless of the timing offset among its rows as shown for the previous $2\times2$ delay-tolerant code.

\section{Performance Evaluation of Delay-Tolerant codes}
In this section, we evaluate the performance of the proposed distributed space-time codes used by the relays in synchronous as well as asynchronous transmission. Recalling the cooperative system model presented in Section \ref{CoopSysModel}, a virtual MIMO scheme is assumed with $M$ transmit antennas (one per relay) and $N_r$ receive antennas. The decoding is performed using the Sphere Decoder as for the perfect codes in conventional MIMO transmission. However in the case of asynchronous relays, the codewords are transmitted over $T+\mathfrak{d}_{\max}$ symbol intervals resulting in rank deficiency of the channel matrix. In order to tackle this problem, the MMSE-DFE preprocessing \cite{Murugan:2006} is required to precede the lattice decoding so that the transformed channel has always full rank.

The performance are represented in terms of codeword error rate CER and bit error rate BER versus signal-to-noise ratio $E_b/N_0$ per receive antenna, which is adjusted as
\begin{equation}
\frac{E_b}{N_0}\Big|_{\mathrm{dB}}=\frac{E_s}{N_0}\Big|_{\mathrm{dB}}-10\log R
\end{equation}
where $E_s$ is the average energy per receive antenna and $R$ is the code rate in bits per channel use (bpcu).

\subsection{Performance Comparison of Existent $2\times 2$ Codes}
For $2\times 2$ schemes, we consider the full-rate full-diversity existent space-time codes in this dimension, namely the Golden code $\mathbf{G}$ \cite{GC}, or its variation matrix $\mathbf{C}$ proposed in \cite{Standard}
\begin{equation*}
\mathbf{C}(\mathbf{s})\triangleq\frac{1}{\sqrt{2(1+r^{2})}}
\left[
  \begin{array}{cc}
    s_1+irs_4 & rs_2+s_3 \\
    s_2-rs_3 & irs_1+s_4 \\
  \end{array}
\right]\, ,\, r=\theta_1-1,
\end{equation*}
the Silver code (Tirkkonen-Hottinen Code) \cite{Tirkkonen:2001, Biglieri:2009} defined by
\begin{eqnarray*}
&\mathbf{X}=\mathbf{X}_A(s_1,s_2)+\mathbf{T}\mathbf{W}\mathbf{X}_B(s_3,s_4)\\
&\mathrm{with}~~\mathbf{X}(s_i,s_j)=\left[                                                                                                      \begin{array}{cc}
 s_i & -s_j^{*} \\
 s_j & s_i^{*} \\
\end{array}
\right], ~~\mathbf{T}=\left[
                        \begin{array}{cc}
                          1 & 0 \\
                          0 & -1 \\
                        \end{array}
                      \right], ~~
\mathbf{W}=\frac{1}{\sqrt{7}}\left[
             \begin{array}{cc}
               1+i & -1+2i \\
               1+2i & 1-i \\
             \end{array}
           \right]
 \end{eqnarray*}
the Sezginer-Sari code \cite{Sezginer:2007} defined by
 \begin{equation*}
\mathbf{S}(\mathbf{s})=\left[
  \begin{array}{cc}
    as_1+bs_3 & -cs^{\ast}_2-ds^{\ast}_4 \\
    as_2+bs_4 & cs^{\ast}_1+ds^{\ast}_3 \\
  \end{array}\right]\,,
\end{equation*}
$$a=c=\frac{1}{\sqrt{2}}\,,\,b=\frac{(1-\sqrt{7})+i(1+\sqrt{7})}{4\sqrt{2}},\,\,d=-ib,$$
the Damen code $\mathbf{D}$ \cite{Damen:2007} defined by
\begin{eqnarray*}
 \mathbf{D}(\mathbf{s}) = \left[\begin{array}{cc}
  x_1 & -x_3 \\
  x_2 & x_4
\end{array}\right]= \left[\begin{array}{cc} a s_1 +b s_2 -c s_3 -d s_4
                                     & -c s_1 -d s_2 -a s_3 -b s_4 \\
                                     -b s_1 + a s_2 + d s_3 -c s_4 &
                                     -d s_1 + c s_2 -b s_3 + a s_4
\end{array}\right]
\end{eqnarray*}
with $a=\frac{1}{\sqrt{(5+\sqrt{5})(2+\sqrt{2})}},~b=\frac{1}{\sqrt{(5-\sqrt{5})(2+\sqrt{2})}},~
c=\frac{1}{\sqrt{(5+\sqrt{5})(2-\sqrt{2})}},~d=\frac{1}{\sqrt{(5-\sqrt{5})(2-\sqrt{2})}}$
and the new proposed code $\bm{\Gamma}$ given in equation (\ref{equ713}). These codes are compared in a distributed setup with and without delays.
Note that the code $\mathbf{D}$ has been proved to verify the NVD criterion for any constellation carved from $\mathbb{Z}[i]$ and to be delay-tolerant \cite{Sarkiss:2008}.

In the above $2\times 2$ schemes, the codewords matrices contain $4$ modulated information symbols carved from $4$-QAM constellation and transmitted over $T=2$ channel uses. The transmission rate is hence  $R=\frac{8}{2+\mathfrak{d}_{\max}}$, where $\mathfrak{d}_{\max}=0, 1$ is the maximum delay with $\mathfrak{d}=(1,0)$ the delay profile in asynchronous transmission.

\begin{figure} [h]%
\centering %
\includegraphics[width = 0.8\textwidth]{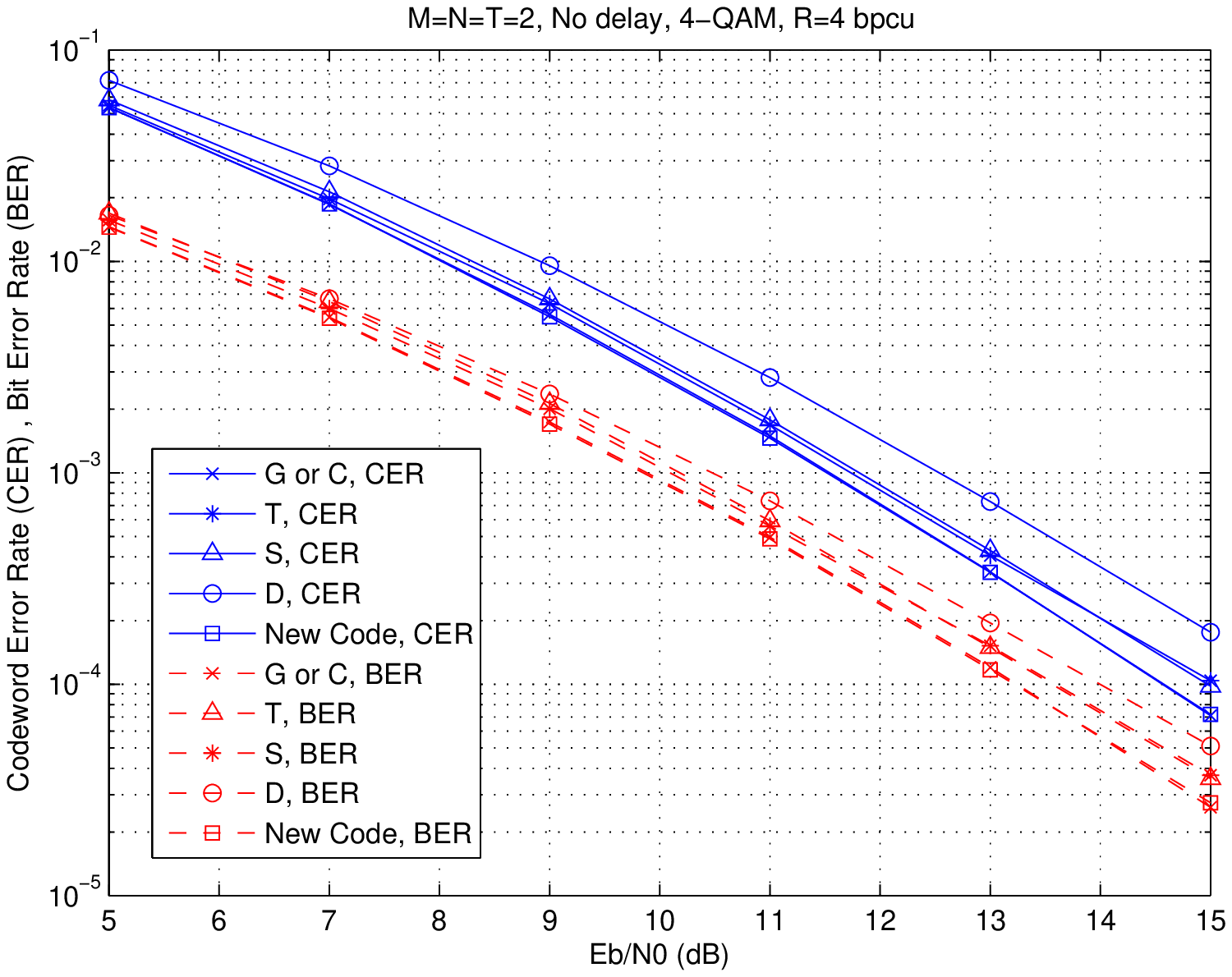} \vspace{-16pt}%
\caption{Performances of the codes $\bm{\Gamma}$, $\mathbf{D}$, $\mathbf{G}$ (or $\mathbf{C}$), $\mathbf{T}$ and $\mathbf{S}$ without delay} %
\label{Code2x2NoDelay}%
\end{figure}

Figure \ref{Code2x2NoDelay} shows the codes performances for synchronous relays $(\mathfrak{d}_{\max}=0)$. Observe that the Golden code $\mathbf{G}$ (or $\mathbf{C}$) outperforms all the other codes. For example, it has about $1$ dB and $0.5$ dB gains over $\mathbf{D}$ and $\mathbf{T}$, $\mathbf{S}$ at a BER of $10^{-4}$, respectively. Note also that the new code $\bm{\Gamma}$ gives the same performance of the Golden code.

\begin{figure}[h]
 \centering
  \includegraphics[width=0.8\columnwidth]{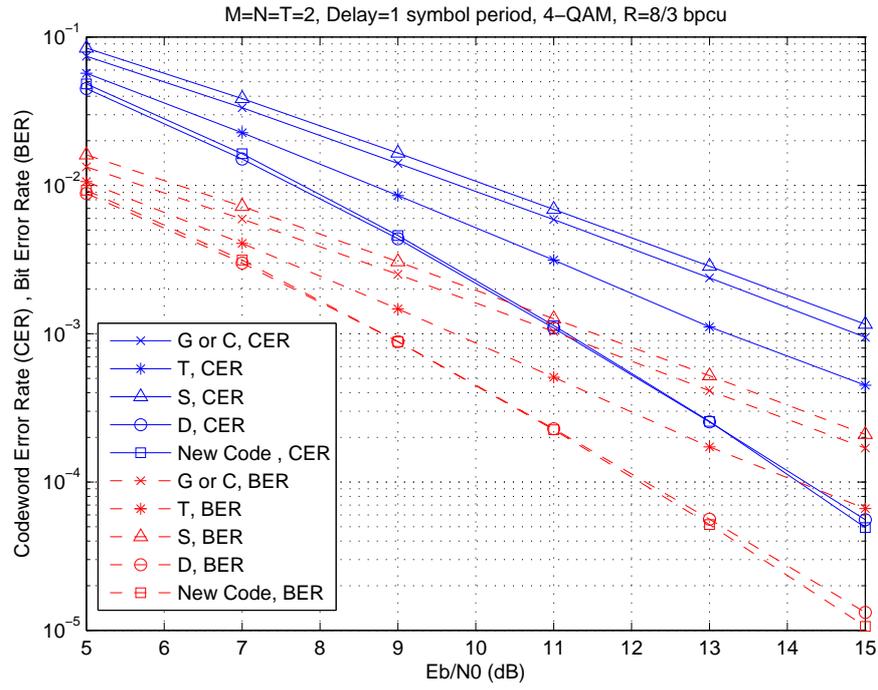}\vspace{-16pt}%
  \caption{Performances of the codes $\bm{\Gamma}$, $\mathbf{D}$, $\mathbf{G}$ (or $\mathbf{C}$), $\mathbf{T}$ and $\mathbf{S}$ with a delay of $1$ symbol period}
\label{Code2x2Delay}
\end{figure}

Whereas for asynchronous relays, the situation is reversed between codes $\mathbf{D}$ and $\mathbf{G}$ for a delay of one symbol period since the latter is not delay-tolerant. It can be seen in Figure \ref{Code2x2Delay} that both delay-tolerant codes $\bm{\Gamma}$ and $\mathbf{D}$ provide gains of $2$ dB and more than $3$ dB over codes $\mathbf{T}$ and $\mathbf{G}$, $\mathbf{S}$ at a BER of $10^{-4}$, respectively. In addition, it can be noticed that $\bm{\Gamma}$ performs almost similar to $\mathbf{D}$,  and it merely improves for high SNR $ (>13$ dB$)$ by $0.2$ dB at a BER of $2\times10^{-5}$.

\begin{figure}[h]
 \centering
  \includegraphics[width=0.8\columnwidth]{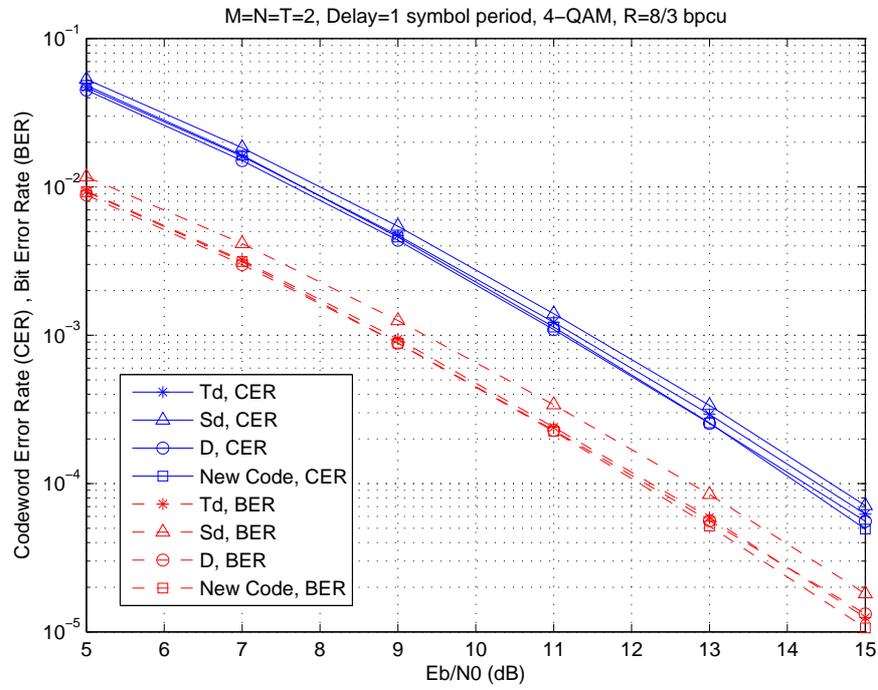}\vspace{-16pt}
  \caption{Performances of the codes $\bm{\Gamma}$, $\mathbf{D}$, $\mathbf{T}_d$ and $\mathbf{S}_d$ with a delay of $1$ symbol period}
\label{Code2x2NewDelay}
 \end{figure}

Using the unitary matrices $\mathbf{U}$ and $\mathbf{V}$ that provide the new code $\bm{\Gamma}$ from the Golden code $\mathbf{G}$, we can also obtain new delay-tolerant codes based on $\mathbf{T}$ and $\mathbf{S}$ codes as
 \begin{equation}
 \mathbf{T}_{d}=\mathbf{U}\mathbf{T}\mathbf{V}\,\,\, \mathrm{and}\,\,\, \mathbf{S}_{d}=\mathbf{U}\mathbf{S}\mathbf{V}
 \end{equation}
Note that $\mathbf{U}$ and $\mathbf{V}$ are not necessarily the optimal matrices for these codes, but they allow to have new delay-tolerant codes with the same determinants as the initial ones. One can easily verify as demonstrated for code $\bm{\Gamma}$ that the product distances associated with these new codes are non-zero over $\mathbb{Z}[i]$. Figure \ref{Code2x2NewDelay} depicts the performances of the new codes for asynchronous relays with a delay of $1$ symbol period. It can be observed that all these delay-tolerant codes preserve their diversity and that the code $\bm{\Gamma}$ gives the best performance. At a BER of $2\times10^{-5}$, it gains about $0.2$ dB and $0.8$ dB over $\mathbf{T}_d$ and $\mathbf{S}_d$, respectively.

\subsection{Performance of $3\times 3$ Codes}
For the $3\times 3$ schemes, $9$ modulated symbols carved from $4$-HEX constellation $(\in\mathbb{Z}[j])$ are transmitted at a rate of $R=\frac{18}{3+\mathfrak{d}_{\max}}$ bpcu, where $\mathfrak{d}_{\max}=0,2$ is the maximum delay and $\mathfrak{d}=(2,1,0)$ the delay profile in asynchronous transmission.

\begin{figure}[h]
 \centering
  \includegraphics[width=0.8\columnwidth]{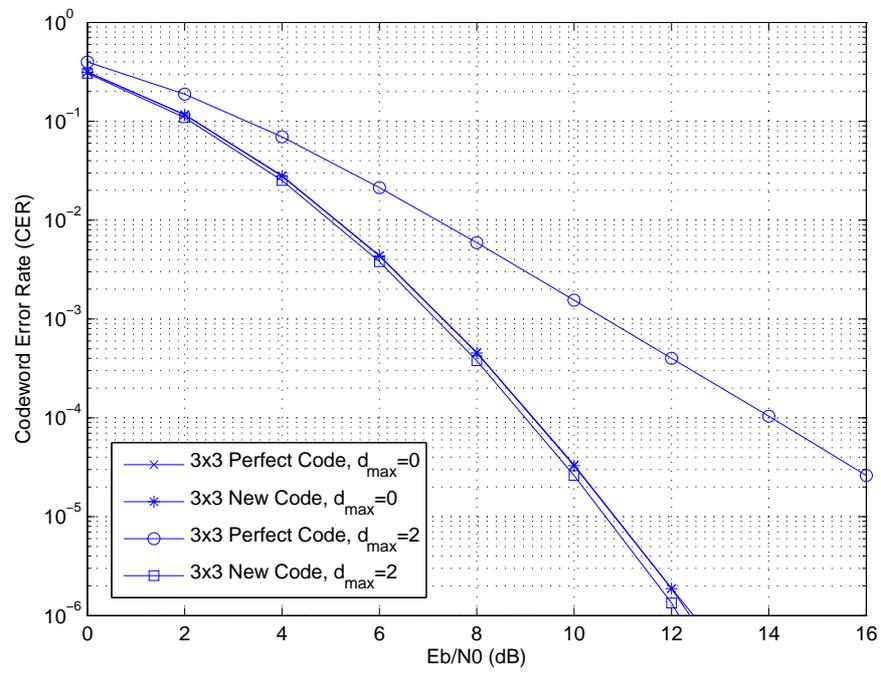}\vspace{-16pt}%
  \caption{Performances of $3\times3$ codes w/wo delay}
\label{Code3x3Delay}
\end{figure}

In Figure \ref{Code3x3Delay}, we can observe that both the perfect code and the new code $\bm{\Gamma}$ have the same performance for synchronous relays. Whereas for asynchronous relays, the delay-tolerant code preserves the diversity and provides a gain of $5$ dB over the $3\times3$ perfect code at CER of $10^{-4}$ for $\mathfrak{d}_{\max}=2$.

\section{Conclusion}
In this paper, we have proposed new delay-tolerant space-time codes based on the perfect codes algebras. Using tensor product of the perfect code field extension with another field extension of the same order $M$ over the same base field and which Galois extensions have coprime discriminants, we build rotated lattices in higher dimension in order to construct $M\times M$ codes. A key parameter in the construction is the coefficients $\phi_l$ that allow to preserve the same properties of the perfect codes in synchronous transmission.

We have found that $\phi_l$ corresponding to the coefficients of the Fourier matrix in dimension $M$ yield the same non-vanishing determinants as the perfect codes. These codes besides having full-rate, full-diversity, uniform energy per transmit antennas ($|\phi_l|=1$) and are information lossless, they have the NVD property and thus are optimal DMT achieving in synchronous case.

In addition, for asynchronous transmission, we have proved for $M=2,3,4$ that the new codes preserve their full-diversity and are delay-tolerant for arbitrary delay profiles. This property is obtained thanks to the non-zero product distances over $\mathbb{Z}[i]$ or $\mathbb{Z}[j]$ and the full-rank minors of the delayed matrices. 

%
%
%
%


\newpage



\end{document}